\pdfoutput=1
\documentclass[twocolumn]{aastex63}
\graphicspath{{figures/}}

\received{September 15, 2020}
\revised{December 18, 2020}
\accepted{to ApJ December 21, 2020}
\shorttitle{Semi-Empirical Models of GJ 832 and GJ 581}
\shortauthors{Tilipman et al.}
\graphicspath{{./}{figures/}}
\begin{document}
	
	\title{Semi-Empirical Modeling of the Atmospheres of the M Dwarf Exoplanet Hosts GJ 832 and GJ 581}

	\author[0000-0001-9361-6629]{Dennis Tilipman}
	\affiliation{National Solar Observatory, Boulder, CO 80309}
	\affiliation{Department of Astrophysical and Planetary Sciences, University of Colorado, Boulder, CO 80309}
	\email{dennis.tilipman@colorado.edu}
	
	\author{Mariela Vieytes}
	\affiliation{Instituto de Astronom\'\i a y F\'\i sica del Espacio (UBA-CONICET), Buenos Aires, Argentina}
	\affiliation{National University of Tres de Febrero, 2736, AHF, Av. Gral. Mosconi, B1674 Senz Pea, Buenos Aires, Argentina}
	
	\author{Jeffrey L. Linsky}
	\affiliation{Department of Astrophysical and Planetary Sciences, University of Colorado, Boulder, CO 80309}
	\affiliation{JILA, University of Colorado, Boulder, CO 80309}
	
	\author{Andrea P. Buccino}
	\affiliation{Instituto de Astronom\'\i a y F\'\i sica del Espacio (UBA-CONICET), Buenos Aires, Argentina}
	\affiliation{Departamento de F\'\i sica, FCEyN Universidad de Buenos Aires, Buenos Aires, Argentina}
	
	\author[0000-0002-1002-3674]{Kevin France}
	\affiliation{Department of Astrophysical and Planetary Sciences, University of Colorado, Boulder, CO 80309}
	\affiliation{Laboratory for Atmospheric and Space Physics, University of Colorado, 600 UCB, Boulder, CO 80309, USA}
	
%	\author{other authors for paper, no other authors for comps 2}
	
	
	
	\begin{abstract}
		
		Stellar ultraviolet (UV) radiation drives photochemistry, and extreme-ultraviolet (EUV) radiation drives mass loss in exoplanet atmospheres. However, the UV flux is partly unobservable due to interstellar absorption, particularly in the EUV range (100--912 \AA). It is therefore necessary to reconstruct the unobservable spectra in order to characterize the radiation environment of exoplanets. In the present work, we use a radiative transfer code \texttt{SSRPM} to build one-dimensional semi-empirical models of two M dwarf exoplanet hosts, GJ 832 and GJ 581, and synthesize their spectra. \texttt{SSRPM} is equipped with an extensive atomic and molecular database and full-NLTE capabilities. We use observations in the visible, ultraviolet, and X-ray ranges to constrain atmospheric structures of the modeled stars. The synthesized integrated EUV fluxes are found to be in good agreement with other reconstruction techniques, but the spectral energy distributions (SEDs) disagree significantly across the EUV range. More than 2/3 of the EUV flux is formed above $10^5$ K. We find that the far ultraviolet (FUV) continuum contributes 42--54\% of the entire FUV flux between 1450--1700 \AA. The comparison of stellar structures of GJ 832 and GJ 581 suggests that GJ 832 is a more magnetically active star, which is corroborated by other activity indicators.
		
	\end{abstract}

	
	
	\section{Introduction} \label{sec:intro}

	M dwarfs are the most common type of stars, particularly among the subset of stars known to host exoplanets. This is due to both their small radii, causing transiting exoplanets to block a larger portion of starlight, and their low masses, which are conducive to radial velocity detections. As the exoplanet sample increases with the recent launch of \textit{TESS} (\textit{Transiting Exoplanet Survey Satellite}) and ground follow-up observations, so does the need for understanding the properties of exoplanet host stars \citep{lit:Linskybook}.
	
	Ultraviolet (UV) and X-ray stellar flux has several crucial effects on exoplanet atmospheres. O$_3$ is photodissociated by near-UV (1700--3000 \AA) radiation, while far-UV (1150--1700 \AA) emission lines, particularly H I Lyman $\alpha$ (Ly$\alpha$, 1215.67 \AA), photodissociate O$_2$, H$_2$O, CO$_2$, CH$_4$, and N$_2$O \citep{lit:2005Segura,lit:2015Miguel,lit:Loyd2016}. High FUV-to-NUV ratios can, therefore, lead to a build-up of abiotic oxygen and ozone \citep{lit:2012Hu,lit:2014Tian,lit:2015Gao,lit:2015Harman}. In primordial atmospheres, the UV flux can trigger haze formation \citep{lit:2006Trainer,lit:2017Arney}. Additionally,  FUV radiation causes atomic hydrogen to accumulate in the outer atmosphere, which can then be ionized by extreme-UV (100--912 \AA) and X-ray ($<$100 \AA) photons. This high energy radiation heats and expands the thermosphere of close-in exoplanets, thereby driving atmospheric escape via ion pick-up and hydrodynamic outflow \citep[e.g.,][]{lit:2009Murray-Clay,lit:2012Owen,lit:2015Tripathi}. This leads to significant ocean loss from terrestrial planets over geological times \citep[and references within]{lit:2015Luger,lit:2007Lammer}. These effects have implications for habitability, since spectral energy distributions (SED) of M dwarfs differ in several important ways from the SEDs of Sun-like stars, as discussed in further sections.
	
	UV observations of even the closest M dwarfs are hindered due to interstellar absorption. Many UV emission lines are partially absorbed by the interstellar medium (ISM) (e.g., Ly$\alpha$, Mg II h \& k doublet at 2796 and 2804 \AA, C II doublet at 1334.5 and 1335.7 \AA, O I line at 1302 \AA), while the FUV continuum has been detected in only a handful of M dwarfs due to their faint signals \citep{lit:2010aHoudebine,lit:Loyd2016,lit:2020Becker}. EUV stellar fluxes are almost completely unavailable due to interstellar hydrogen absorption and a lack of observing facilities operating in that wavelength range. Low sensitivity spectra of ten M dwarfs have been observed in the 90--360 \AA\ range by \textit{EUVE} (\textit{Extreme Ultraviolet Explorer}) \citep{lit:EUVE}, but there have been no more sensitive instruments in space to observe M dwarfs at these wavelengths.
	
	In the absence of direct observations, missing parts of spectra can be reconstructed either by fitting observed line profiles for interstellar absorption and subsequently removing said absorption, or by employing empirical scaling relations. \citet{lit:2005Wood} have reconstructed 33 Ly$\alpha$ profiles from spectra in the HST archives, and \citet{lit:Youngblood2016} reconstructed intrinsic Ly$\alpha$ profiles for 11 stars in the MUSCLES (Measurements of the UV Spectral Characteristics of Low Mass Exoplanetary Systems) survey. They also observed a correlation between Ly$\alpha$ and Mg II surface fluxes. More recently, \citet{lit:2019Schneider} reconstructed two new Ly$\alpha$ profiles and updated empirical relations between Ly$\alpha$ and FUV fluxes. Since EUV flux is formed at a range of heights extending from the upper chromosphere to the lower corona, it can be constrained from below using chromospheric UV features \citep{lit:LinskyLyalpha,lit:Youngblood2017,lit:2018France} or from above using coronal X-ray emission \citep{lit:2011Sanz,lit:2015Chadney}. Duvvuri et al. (submitted) reconstructed stellar EUV fluxes using the differential emission measure (DEM) approach. They make use of a multitude of observable FUV and X-ray features.
	
	There have been extensive efforts to compute synthetic spectra by building models of stellar atmospheres. \texttt{PHOENIX} photosphere models, including those computed in a 3-D framework, have been successful in reproducing infrared (IR) and visible spectra for a wide range of stars \citep[e.g.,][]{lit:2010Hauschildt}. However, they do not typically account for chromospheric and coronal non-thermal heating and, therefore, underestimate flux at short wavelengths and fail to reproduce profiles of the lines that are proxies for magnetic activity. Such lines include Ca II H \& K doublet, which scales directly with chromospheric nonradiative heating \citep{lit:2009Walkowicz,lit:2011Gomes}, Na I D doublet at 5890 and 5896 \AA, and H$\alpha$ line at 6563 \AA, which are all reliable activity indicators for most active M dwarfs \citep{lit:2007Diaz,lit:2007Cincunegui,lit:2009Walkowicz,lit:2009Houdebineetal}. Many of the previous chromospheric models have been built to fit only those and several other lines \citep[e.g.,][]{lit:2005Fuhrmeister,lit:2009Houdebine,lit:2010aHoudebine}. \citet{lit:2011Sanz} synthesized EUV spectra for 82 F to M dwarfs, including GJ 832, from their coronal models. They used X-ray data from \textit{ROSAT} \textit{(ROentgenSATellit)}, \textit{XMM-Newton} \textit{(X-ray Multimirror Mission)}, and \textit{Chandra} to constrain the thermal structure of the coronae, but their models did not include chromosphere-to-corona transition regions (TR) where much of the EUV radiation is also formed. \citet{lit:Peacock} used improved \texttt{PHOENIX} models with partial redistribution (PRD) spectral line formation capabilities to synthesize EUV-to-IR spectra of the ultracool dwarf TRAPPIST-1, but their models do not include the corona, and many of the EUV lines were computed assuming local thermodynamic equilibrium (LTE). Shortly thereafter, \citet{lit:Peacock832} further improved upon these \texttt{PHOENIX} models by extending PRD formalism to calculations of the chromospheric Mg II h \& k and Ca II H \& K lines. They synthesized EUV-to-IR spectra of three M dwarfs, including GJ 832, but once again, the models extended only to 200,000 K. \citet{lit:Fontenla832} synthesized full panchromatic spectra of the M dwarf GJ 832 using the Solar-Stellar Radiation Physical Modeling (\texttt{SSRPM}) software, which is the basis of the present work and is described in the next section. Their stellar atmosphere model extends from the photosphere to the corona; however, the visible spectra used in that work were obtained using a faulty calibration procedure described in Section \ref{sec:spectra_vis}. 
	
	The goal of this paper is to revise the model of the M2 V dwarf GJ 832 using recalibrated visible spectra, as well as to build a model for GJ 581 (M3 V). This is done in order to synthesize and characterize the unobservable parts of spectra, and to study the physical structure of the model atmospheres. These stars were chosen because 1) they are both exoplanet hosts, 2) exquisite UV spectral data are available through the MUSCLES web-site for both of them, and 3) they are of similar spectral types, allowing us to make quantitative comparisons between them. The physical parameters of these stars are listed in Table \ref{tab:stellarparams}. 
	
	The structure of this paper is as follows. In Section \ref{sec:methods}, we describe the procedure for computing synthetic spectra (a more detailed description is provided in the Appendix). In Section \ref{sec:spectra}, we compare synthetic spectra with the observations. Unobservable spectra, scaling relations, physical properties of modeled atmospheres, and current challenges are discussed in Section \ref{sec:disc}. We summarize the most important results in Section \ref{sec:outro}.

\begin{deluxetable}{lchccc}
	\tablecaption{Stellar Parameters\label{tab:stellarparams}}
	\tablewidth{0.5\textwidth}
	%\tabletypesize
	\tablehead{
		\colhead{} & \colhead{GJ 832} & \nocolhead{} & \colhead{GJ 581}
	}
	%\decimalcolnumbers
	\startdata
	$d$ (pc) $^1$ & $4.965\pm0.001$ &  & $6.30\pm0.01$  \\
	R$_*$ (R$_{\odot}$) & $0.499\pm0.017^2$ &  & $0.299\pm0.010^3$ \\
	log [Fe/H]$^*$ & $0.06\pm0.04^4$ &  & $-0.33\pm0.12^5$  \\
	log $g$ (cgs) & $4.7^6$ &  & $4.92\pm0.10^5$  \\
	Spectral type & M2$^7$ &  & M3$^8$  \\
	T$_{\text{eff}}$ (K) & $3590\pm100^7$ &  & $3498\pm56^2$  \\
	Age (Gyr)$^{**}$ & $8.4^9$ &  & $4.1\pm0.3^{10}$  \\
	$\log{R'_{HK}}$ $^{11}$ & $-5.1$ &  & $-5.7$  \\
	\enddata
	\tablerefs{(1) \citet{lit:gaia}, (2) \citet{lit:GJ581params}, (3) \citet{lit:GJ876params}, (4) \citet{lit:GJ832metal}, (5) \citet{lit:GJ581and876params}, (6) \citet{lit:GJ832_logg}, (7) \citet{lit:2016Houdebine}, (8) \citet{lit:CARMENES}, (9) \citet{lit:2009Bryden}, (10) \citet{lit:2017Yee}, (11) \citet{lit:2020Melbourne} \\
		$*$ -- solar metallicity was assumed since log [Fe/H] is within 3$\sigma$ of 0.0 for both stars \\
		$**$ -- the ages of both of these stars are poorly constrained, and we caution against relying on any one estimate}
	%	\tablecomments{}
	\vspace{-20pt}
	
\end{deluxetable}

	\section{Modeling Stellar Atmospheres and Computing Synthetic Spectra} \label{sec:methods}	
	
	We build semi-empirical 1-D models of stellar atmospheres using the \texttt{SSRPM} model atmosphere code \citep{lit:Fontenla832}. \texttt{SSRPM} is a modified version of Solar Radiation Physical Modeling (\texttt{SRPM}) \citep[see][and references within]{lit:SRPM2015} in which key differences between the Sun and M dwarfs have been taken into account. Specifically, the \texttt{SSRPM} databases include twenty diatomic molecules commonly detected in M dwarf atmospheres, along with over 2,000,000 molecular lines calculated in LTE. This is in addition to 195 highly ionized species with abundances computed in the effectively optically thin non-LTE (NLTE) approximation, and 55 atoms and ions, including H$_2$, H$^-$, and H, that are computed in full-NLTE, i.e. in the optically thick regime. \texttt{SSRPM} is equipped with PRD capabilities, and we use them to compute transition lines associated with key atomic species, such as H I, Mg II, and Ca II. Overall, the database includes 19,132 atomic levels and 436,045 corresponding transitions. The atomic and molecular data were adapted from CHIANTI 7.1 \citep{lit:CHIANTI}, NIST \citep{lit:NIST}, and TOPbase \citep{lit:TOPbase} databases, along with the TiO molecular line list from \citet{lit:1998Plez}.  While our approach resembles that of \texttt{PHOENIX} or FAL \citep[e.g.,][]{lit:FAL1993,lit:FAL2002} models, the extent of our database and the NLTE treatment of all non-molecular spectral features in our models are both key distinctions. Additionally, we model all atmospheric regions from the photosphere to the corona and include the effects of ambipolar diffusion in our calculations, which are particularly relevant for hydrogen species.
	
	We begin constructing a model of a stellar atmosphere by adopting an appropriate abundance set of elements for the star. We use solar abundances for both stars, since their metallicities do not significantly ($>3\sigma$) deviate from solar [Fe/H] (see Table \ref{tab:stellarparams}). The normalized elemental abundances are shown in Table \ref{tab:abunslvls} of the Appendix, where we discuss the structure of \texttt{SSRPM} in more detail. We assume that relative elemental abundances remain constant throughout the atmosphere. The models are computed on a height grid, with any two adjacent heights separated by no more than half of the pressure scale height. We therefore rely on estimates of surface gravity found in literature (see Table \ref{tab:stellarparams} for references). Heights are fixed, and each height has a set of corresponding parameters: temperature ($T$), microturbulent velocity ($v_t$), and number densities of protons ($n_p$), electrons ($n_e$), and neutral hydrogen atoms ($n_a$).
	
	Unlike grids of stellar models that are constrained solely by theoretical predictions (e.g., \texttt{PHOENIX} and \texttt{MARCS} \citep{lit:MARCS}), our semi-empirical models rely on observational input. We model an atmospheric profile from the bottom up using various spectral features as diagnostics to constrain it. The photosphere structure is typically close to an existing \texttt{PHOENIX} model of a star with similar surface gravity, metallicity, and effective temperature. Following the semi-empirical approach, we then modify the outer atmosphere parameters to match UV and X-ray spectral features and fluxes, along with Ca II H \& K, H$\alpha$, and Na I D lines. We employ our understanding of stellar atmospheres to obtain an initial guess for how temperature and number densities change with height. The overall shapes of our T-P profiles generally resemble those described in existing literature \citep[e.g., ][]{lit:1981Vernazza,lit:Fontenla832,lit:Peacock}. The photosphere is assumed to display low departures from LTE and gradual outward decrease in temperature $T \propto (\tau+2/3)^{1/4}$, where $\tau$ is the optical depth in the visible range. Above the photosphere, the plasma density is lower and the temperature increases steeply in the lower chromosphere due to nonradiative heating. The temperature profile then flattens out in the upper chromosphere. The transition region between the upper chromosphere and the corona (hereafter TR) shows another steep increase in temperature from around $10^4$ to $10^6$ K (Figure \ref{fig:models}). The extent of chromospheric plateau determines the geometrical thickness of the chromosphere and the pressure at which TR occurs, which correlate with emission strength of chromospheric and TR lines, respectively. It must be noted that the chromospheric structure in our models deviates from that described in \citet{lit:Peacock832}, where the plateau in the upper chromosphere is not present. Their models are also distinct from ours in that they adopt three free parameters that set the entire structure of the chromosphere and TR, whereas in our models, each individual point can be changed independently from others. This allows us to fine tune our models to fit individual lines better, but it introduces vastly more degrees of freedom. Finally, the corona is characterized by significantly higher temperatures and lower pressures compared to the chromosphere.

	\section{Comparison between Models and Observations} \label{sec:spectra}
	
	\begin{figure*}
		\gridline{\fig{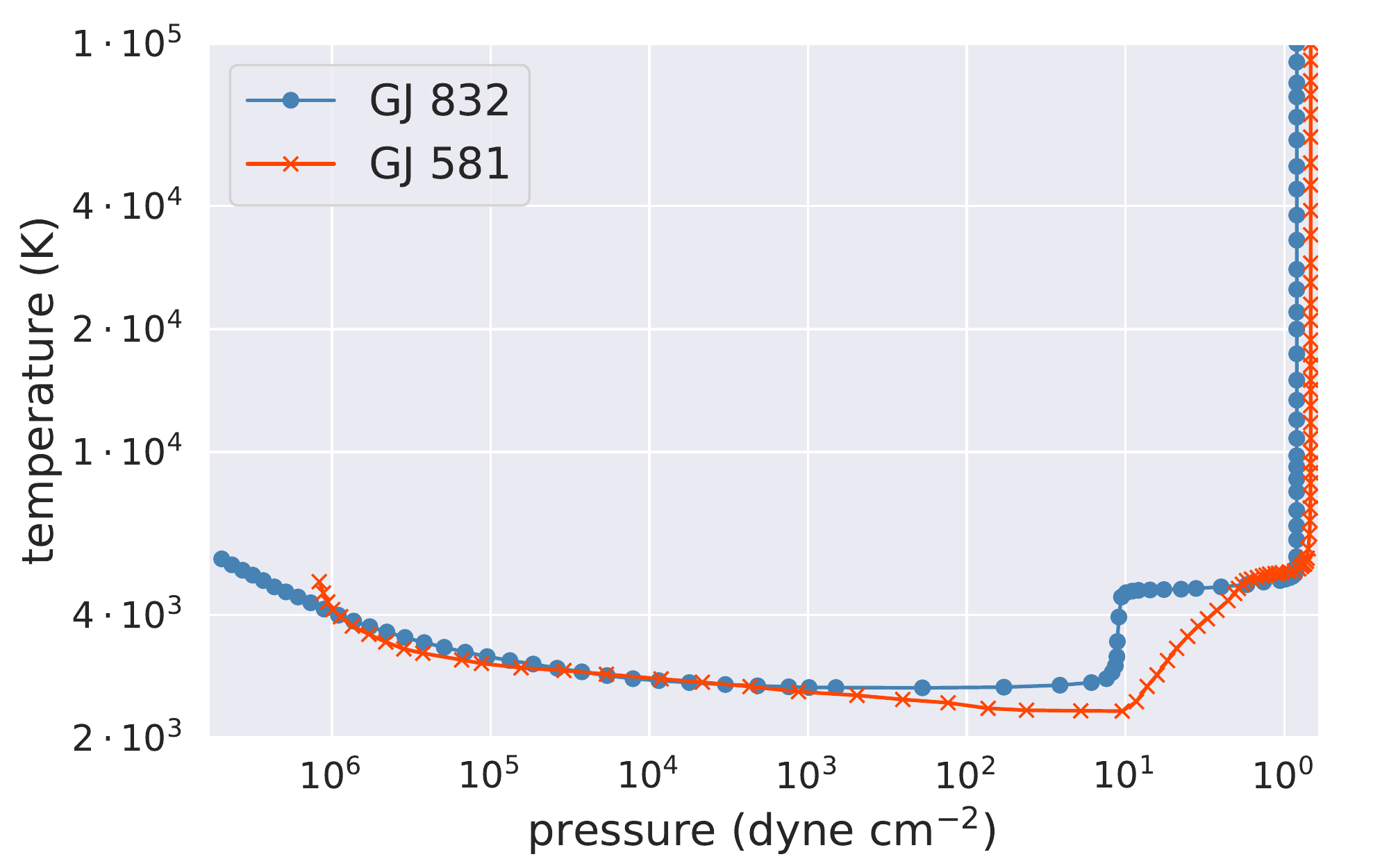}{0.5\textwidth}{}
			\fig{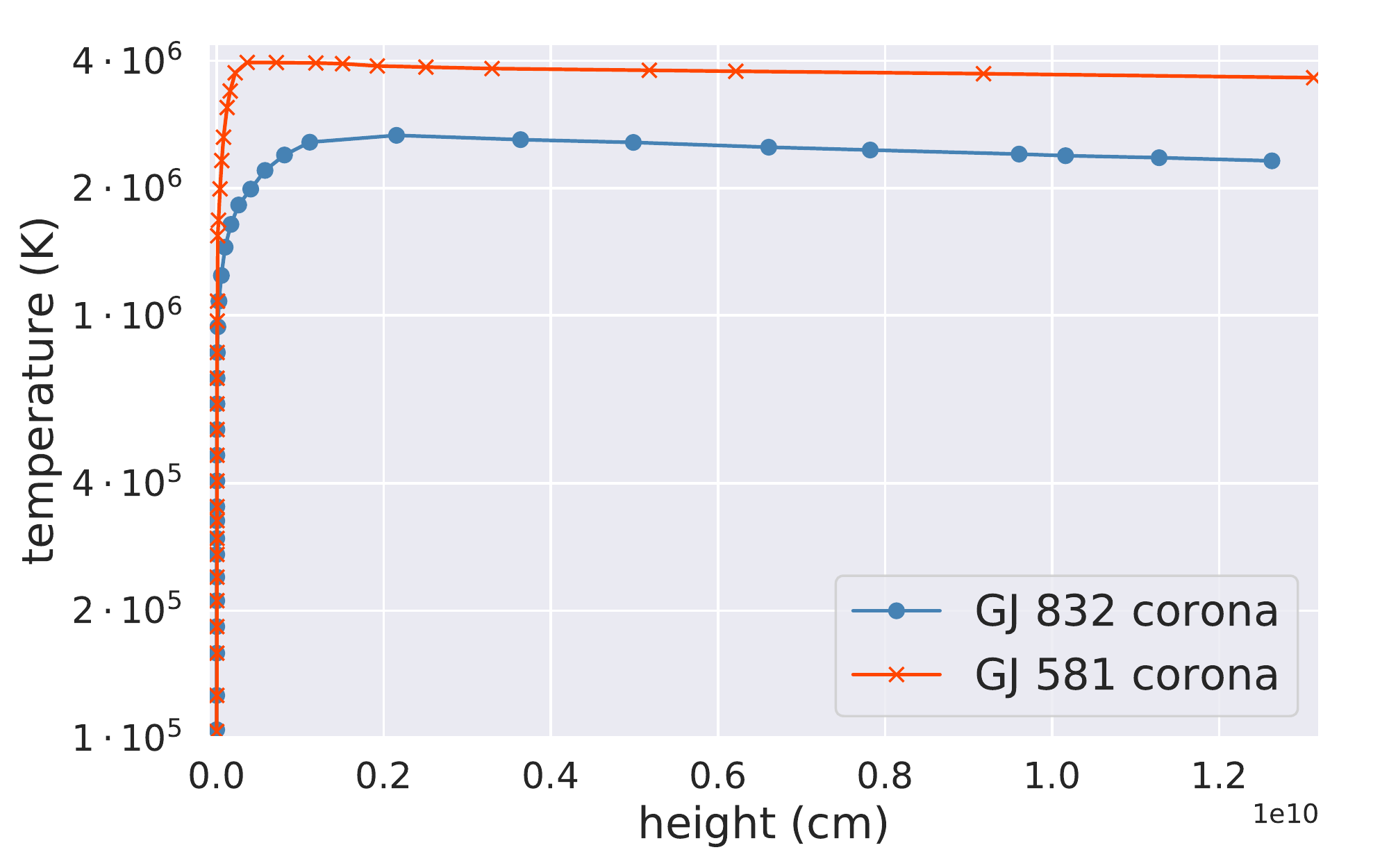}{0.5\textwidth}{}
			\vspace*{-20pt}
		}
		\caption{ Lower (left) and upper (right) model components used to synthesize spectra. For the upper component, temperature is plotted against height because pressure is nearly constant throughout the corona. Points indicate grid spacing.
			\label{fig:models}}
		\vspace*{5pt}
	\end{figure*}
	
	The models used in the present work to synthesize spectra are shown in Figure \ref{fig:models}. The photospheres of the two stars extend from the lowest altitudes until the temperature minimums. Both photospheres are characterized by gradual decline in temperature and are quantitavely similar, except the temperature minimum in the atmosphere of GJ 832 occurs at 2650 K, which is about 200 K hotter than in GJ 581. The temperature gradient in the lower chromosphere of GJ 832 is much steeper than that of GJ 581. The upper boundary of the first steep chromospheric rise marks the lowest points of the upper chromospheres. The upper chromospheric plateaus are at roughly the same temperatures of about 5,000 K, but the TR of GJ 581 occurs at a lower pressure. Finally, the shape of the coronae are similar, but GJ 581 has a hotter corona in our models. The importance of these parameters is discussed in the following sections.
	
	\subsection{Visible Spectrum} \label{sec:spectra_vis}
	
	\begin{figure}
		\gridline{\fig{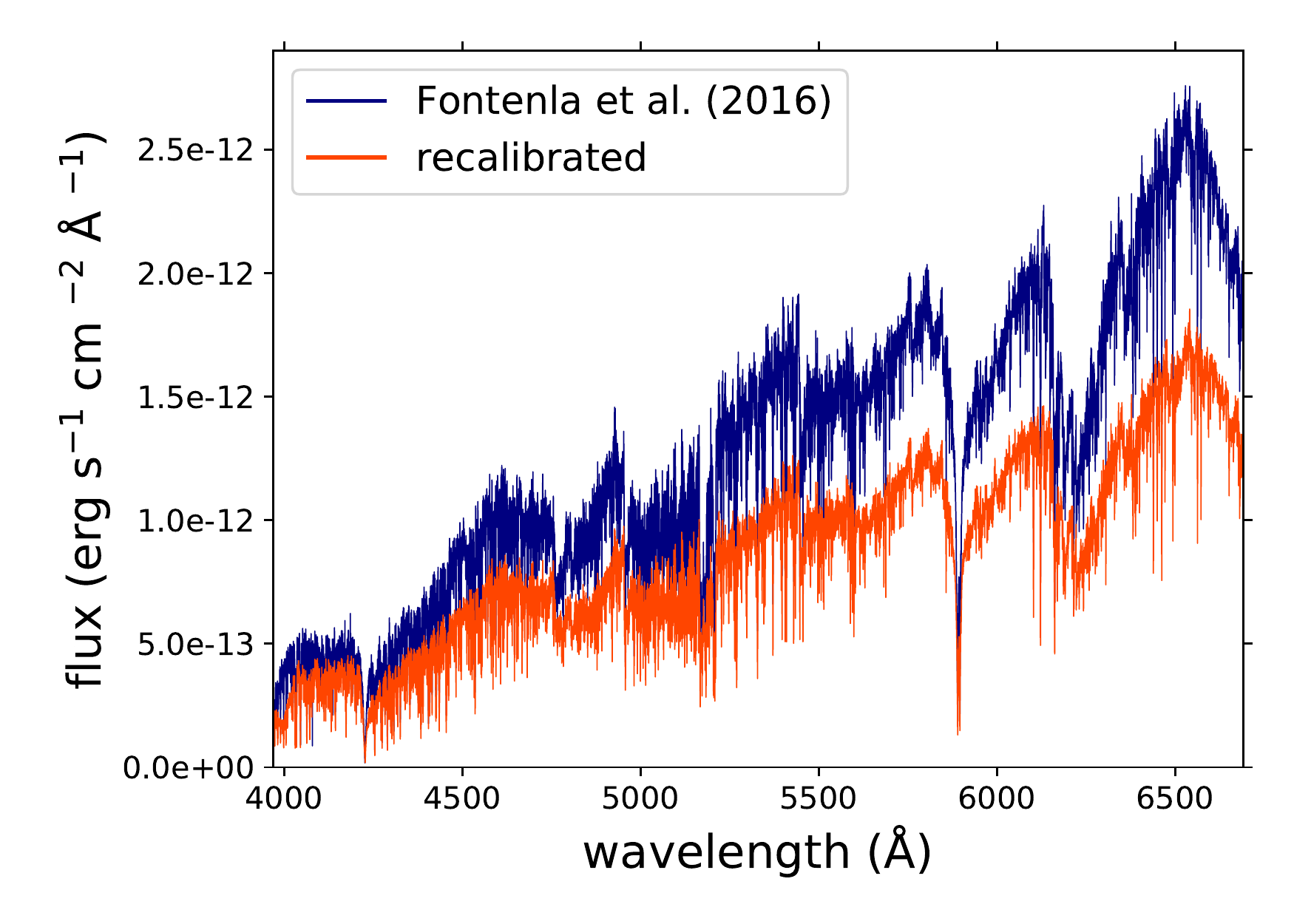}{0.5\textwidth}{}
			\vspace*{-20pt}
		}
		\caption{Echelle spectra of GJ 832 calibrated against the long-slit spectra of GJ 1 (blue) and GJ 832 (red).
			\label{fig:832oldnew}}
	\end{figure}
	
	The visible spectra were obtained with a REOSC spectrograph with the spectral resolution $R$=13,000 on the 2.15 m Jorge Sahade telescope of the Complejo Astron\'{o}mico El Leoncito (CASLEO). We follow the calibration procedure described in \citet{lit:2004Cincunegui}. The echelle spectrum of the target star is calibrated in flux with the long-slit spectrum of the same star. The spectra of GJ 832 and GJ 581 were obtained in September of 2012 and March of 2016, respectively. Each corresponding long-slit spectrum was obtained with the REOSC spectrograph in DS configuration on October, 2014 and July 2014 respectively. Following \citet{lit:2004Cincunegui}, we estimated a 10\% error in the absolute flux in each echelle spectrum. Previously, \citet{lit:Fontenla832} used the long-slit spectra of GJ 1, an M1 dwarf, to calibrate the echelle spectra of GJ 832 in flux, arguing that since the stars are of similar spectral types, the associated calibration error may be around 10\%. They noted, however, that it may be larger, since the differences between GJ 1 and GJ 832 were not precisely known. Indeed, this procedure resulted in a significant overestimate of the flux in the visible range (see Figure \ref{fig:832oldnew}), prompting us to rebuild the model of GJ 832.
	
	\begin{figure*}
		\gridline{\fig{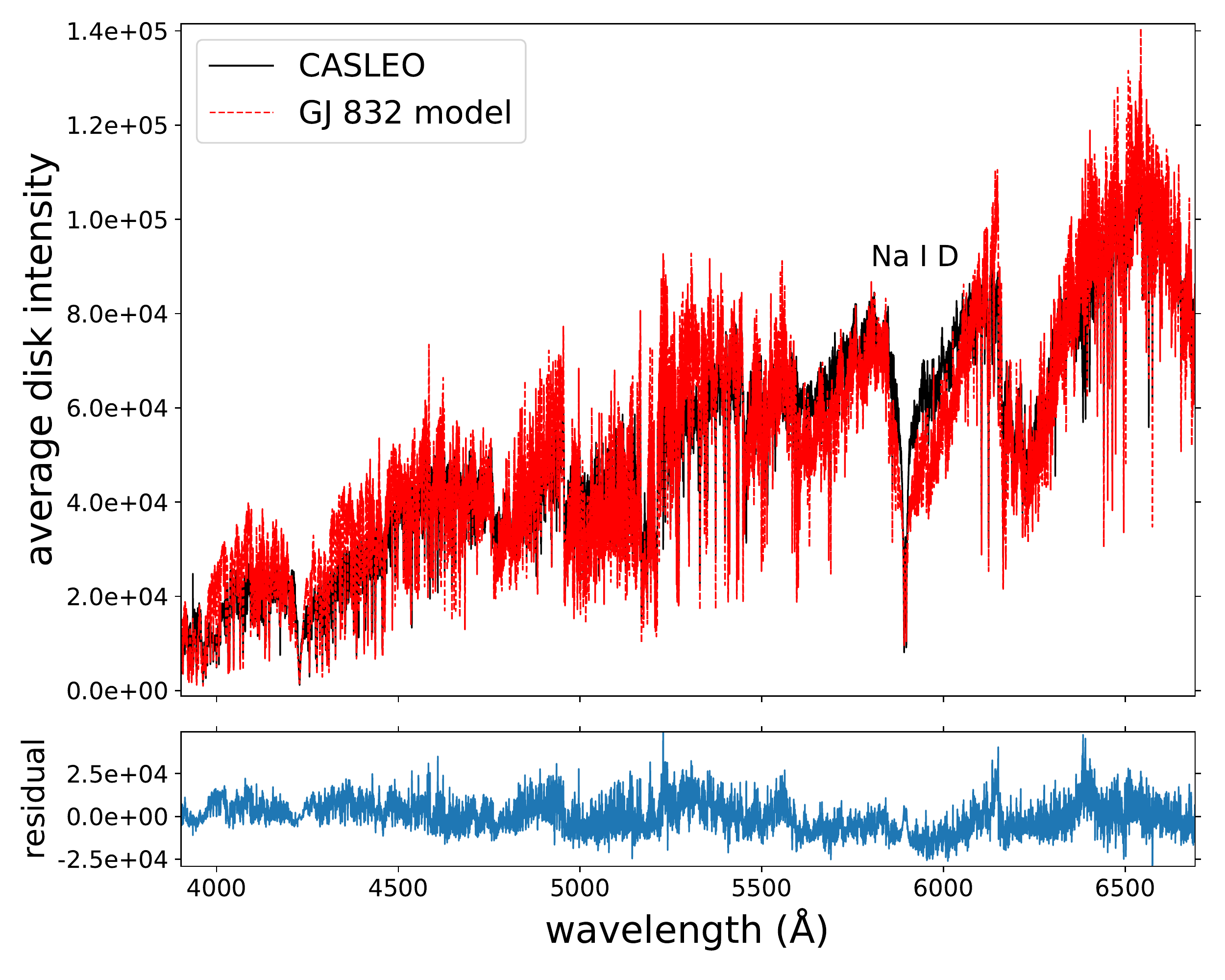}{0.45\textwidth}{}
			\fig{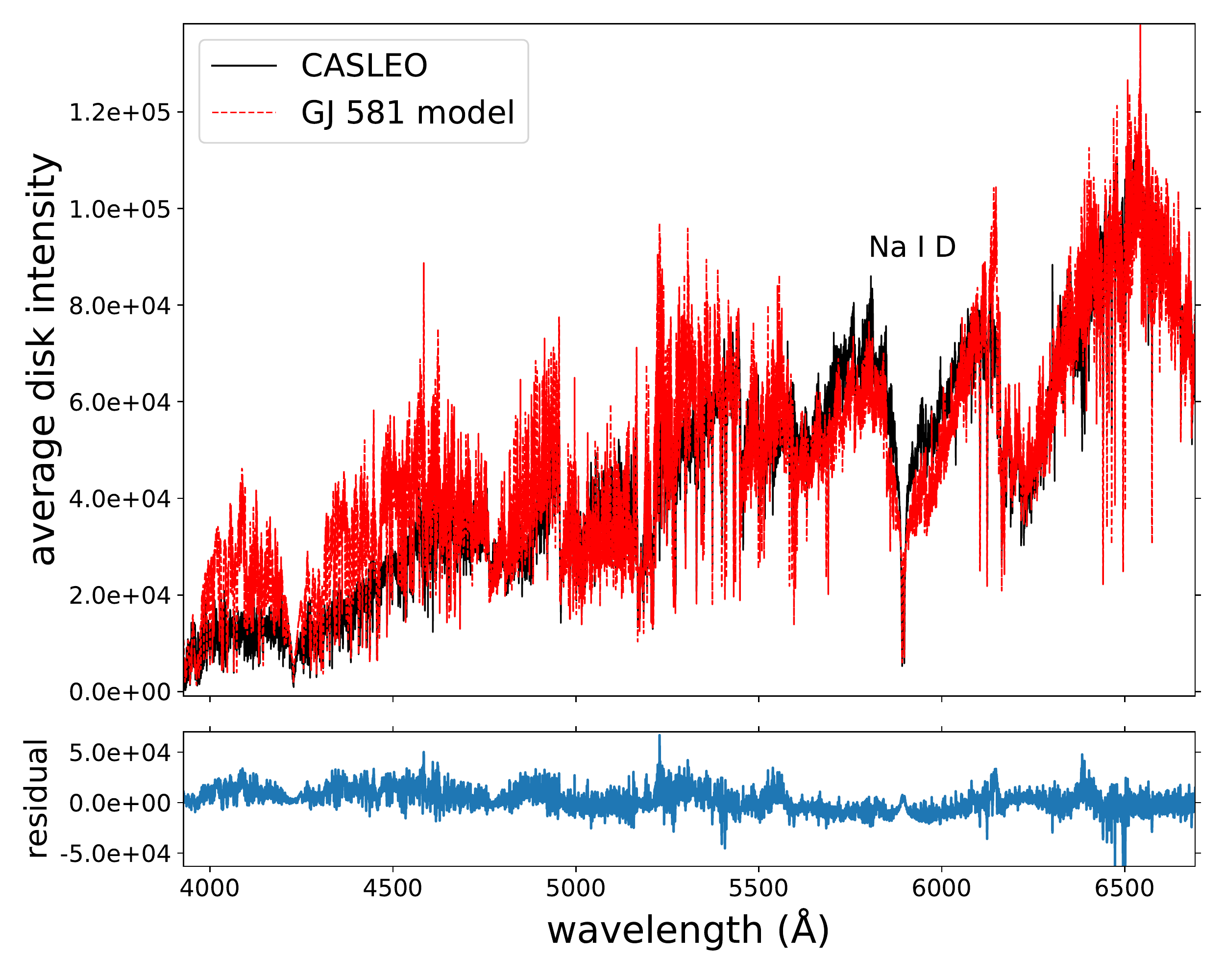}{0.45\textwidth}{}
			\vspace*{-20pt}
		}
		\gridline{\fig{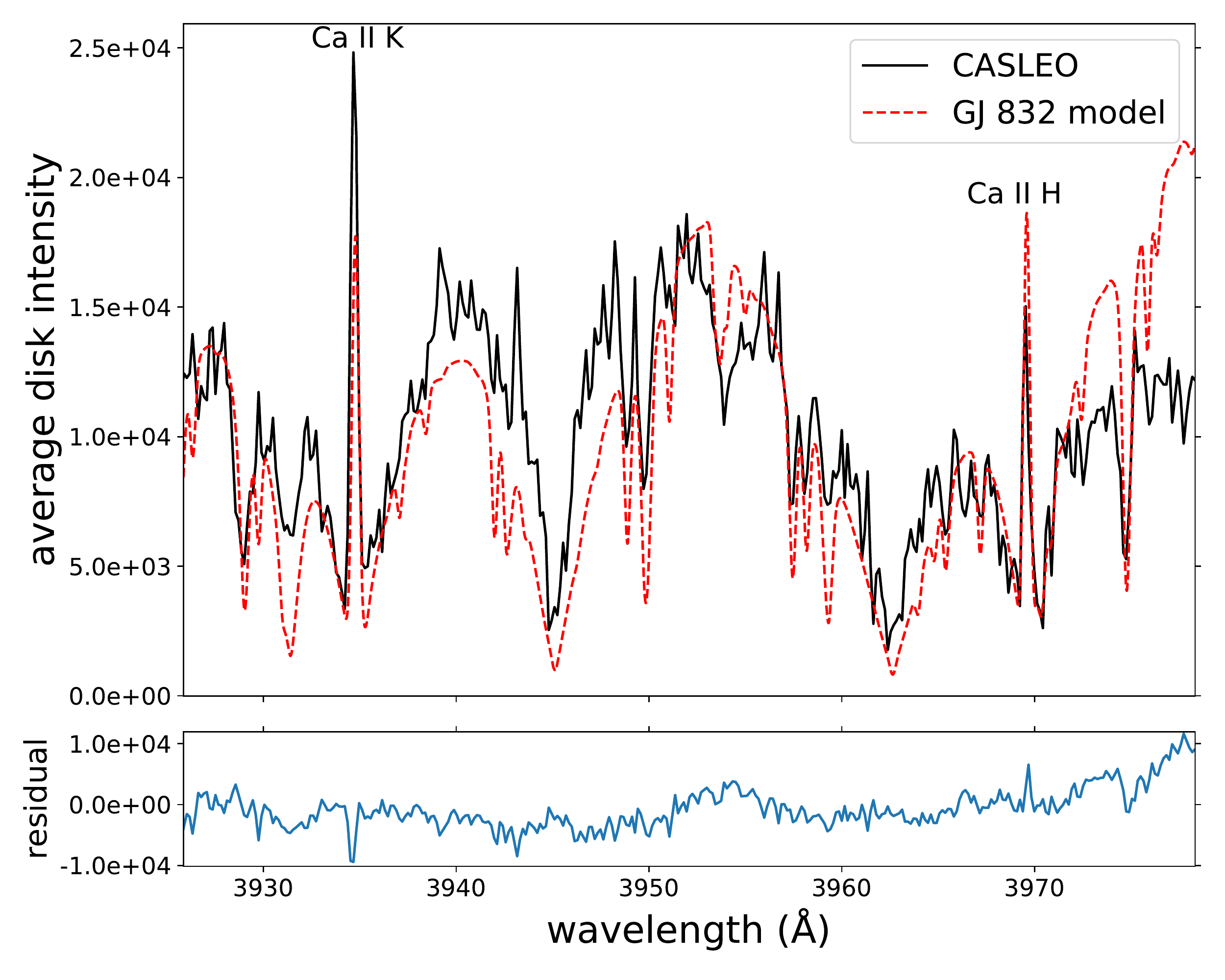}{0.45\textwidth}{}
			\fig{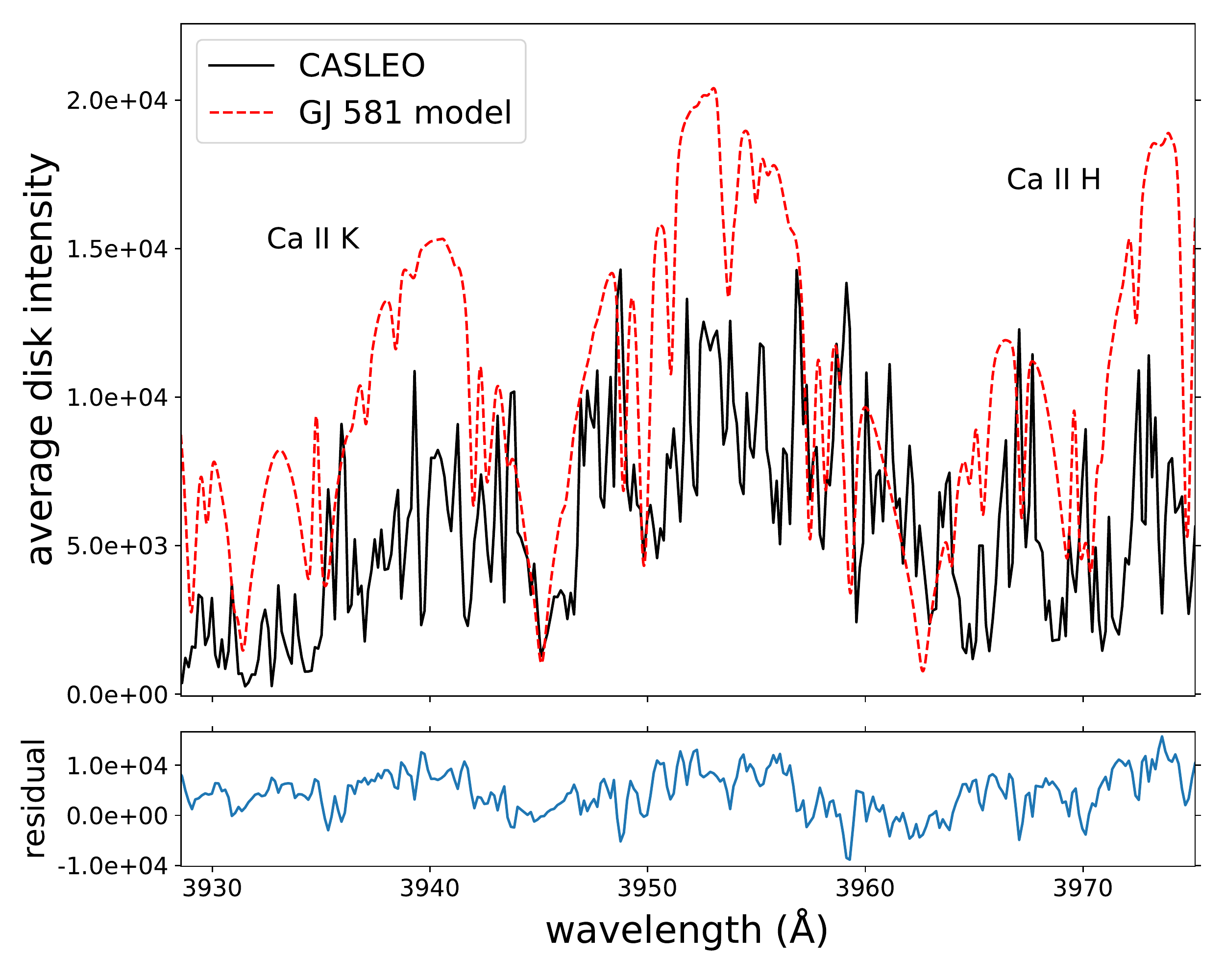}{0.45\textwidth}{}
			\vspace*{-20pt}
		}
		\gridline{\fig{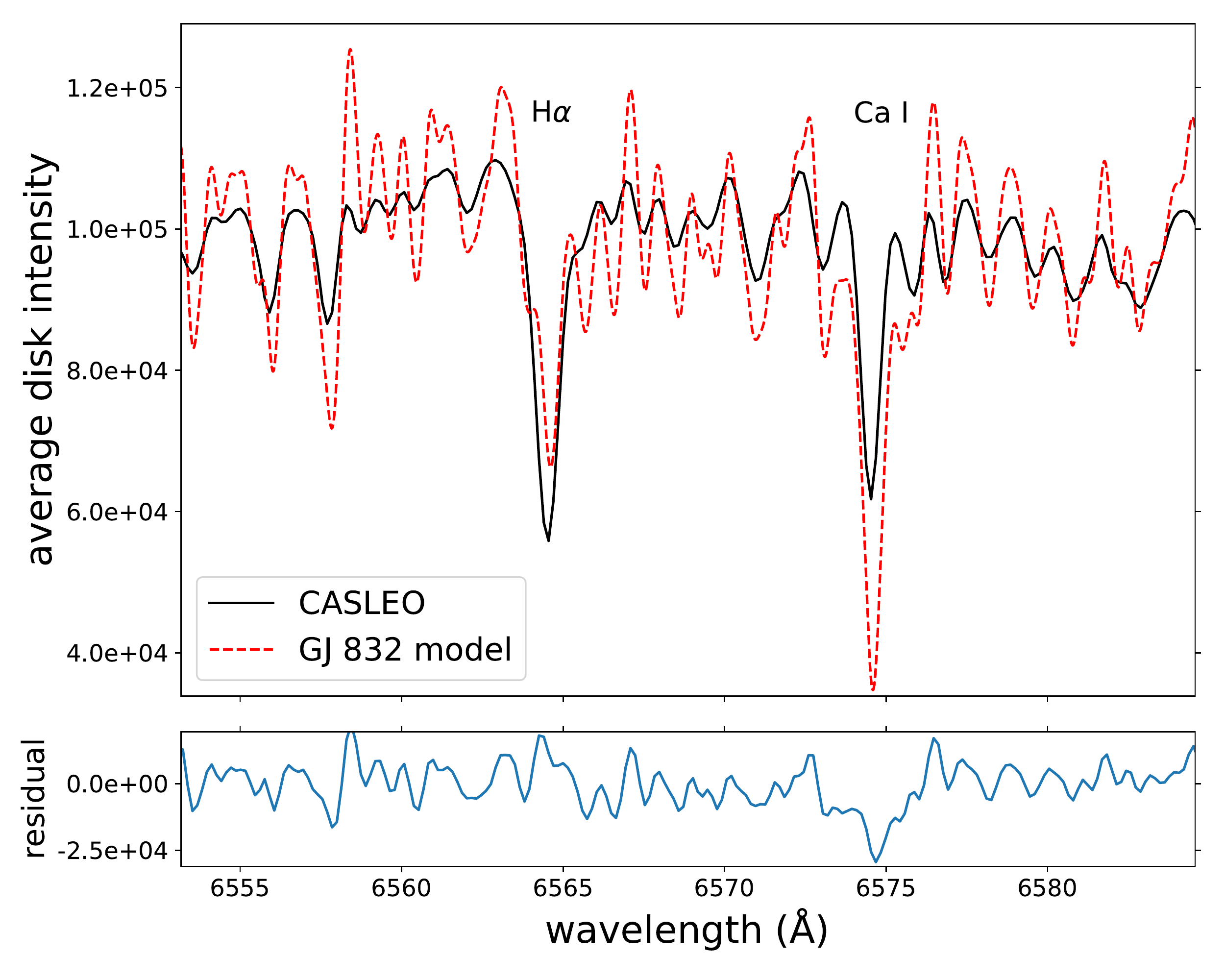}{0.45\textwidth}{}
			\fig{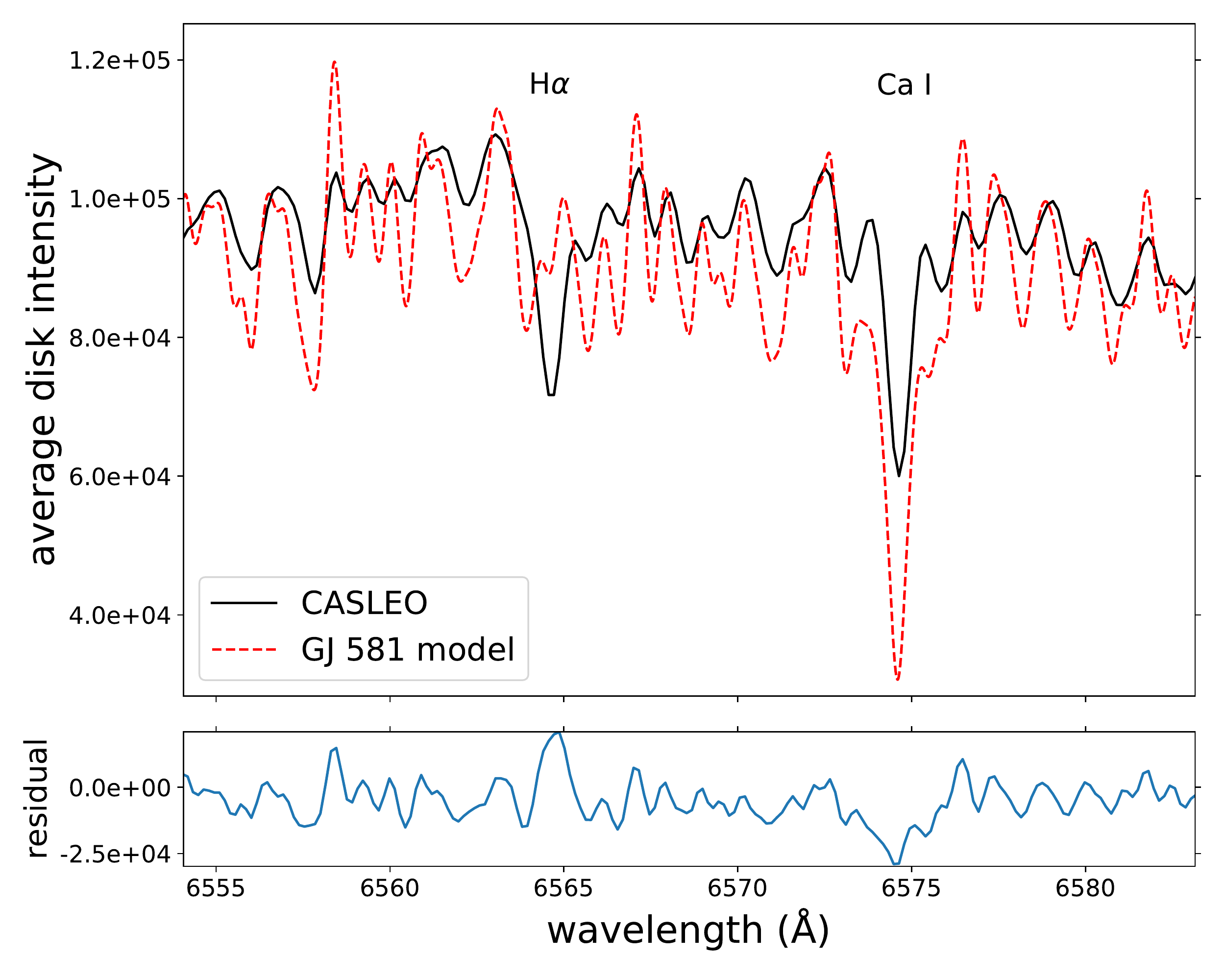}{0.45\textwidth}{}
			\vspace*{-20pt}}
		\caption{Comparison between observed (black) and computed (red) average disk intensities (erg cm$^{-2}$\AA$^{-1}$s$^{-1}$sr$^{-1}$) of GJ 832 (left column) and GJ 581 (right column) in the optical range. \textit{Top row}: the entire range; \textit{middle row}: Ca II H \& K lines; \textit{bottom row}: H$\alpha$ line. The bottom subplots in each panel show the differences between the synthetic and observed spectra for each wavelength. Most important spectral features are highlighted.
			\label{fig:opt}}
	\end{figure*}

	\begin{figure}
	\gridline{\fig{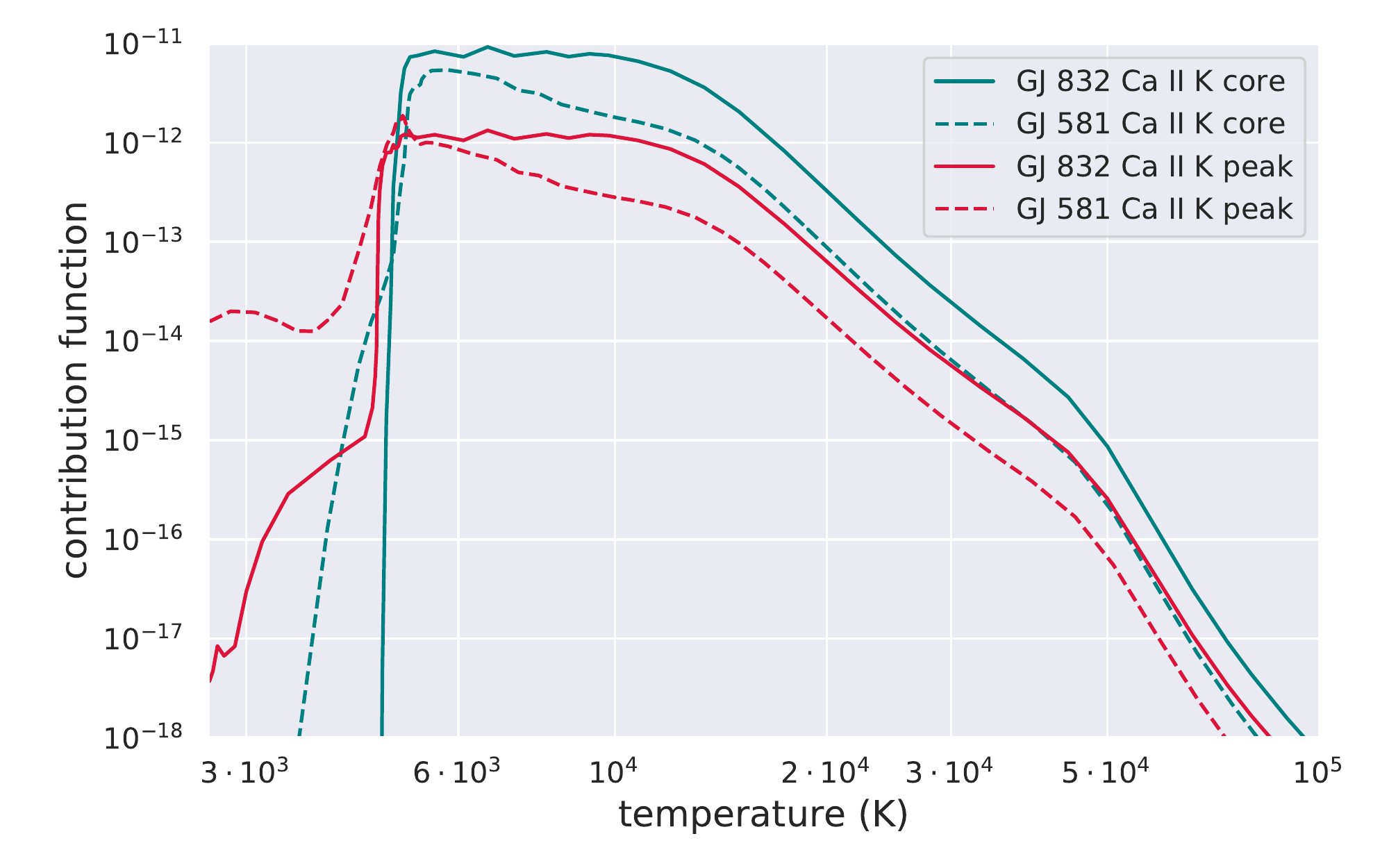}{0.45\textwidth}{}
		\vspace*{-20pt}
	}
	\caption{Contribution functions of Ca II K peak (red) and core (teal). Solid (dashed) lines represent contribution functions in the model of GJ 832 (GJ 581).
		\label{fig:contrib_opt}}
	\end{figure}
	
	\begin{deluxetable*}{lcccccc}
		\tablecaption{Average disk intensities (erg cm$^{-2}$\AA$^{-1}$s$^{-1}$sr$^{-1}$) of key spectral lines \label{tab:lines}}
		%\tablewidth{0pt}
		%\tabletypesize
		\tablehead{
		    %\vspace{-5pt}
			\colhead{Feature \&} & \multicolumn{3}{c}{GJ 832} & \multicolumn{3}{c}{GJ 581            } \\
			\colhead{$\lambda$ (\AA)} & \colhead{synthetic} & \colhead{observed} & error\tablenotemark{\footnotesize{$\mathrm{a}$}} (\%) & \colhead{synthetic} & \colhead{observed} & error (\%)
		}
		%\decimalcolnumbers
		\startdata
		Ca II H \& K 3950 & 1.13E4 &  1.23E4 & 8.13 & 5.21E3 & ... & ...   \\
		Mg II h \& k 2800 & 2.73E4 &  2.17E4\tablenotemark{\footnotesize{$\mathrm{b}$}} & 25.8 & 6.83E3 & 7.19E3\tablenotemark{\footnotesize{$\mathrm{b}$}} & 5.01  \\
		C II 1334 & 84.3 &  133\tablenotemark{\footnotesize{$\mathrm{b}$}} & 36.8 & 43.9 & 62.2\tablenotemark{\footnotesize{$\mathrm{b}$}} & 29.4 \\
		C II 1336 & 132 &  180 & 26.7 & 70.5 &  129 & 45.3 \\
		Si IV 1394 & 125 & 137 & 8.09 & 52.6 & 76.5 & 31.2   \\
		Si IV 1402 & 65.8 & 73.4 & 10.4 & 27.4 & 29.6 & 7.43  \\
		C IV 1548 & 201 & 323 & 37.8 & 325 & 297 & 9.43  \\
		C IV 1551 & 102 & 154 & 33.8 & 163 & 148 & 10.1  \\
		O IV 1401 & 7.92 &  11.6 & 31.7 & 2.38 & 7.37 & 67.7   \\
		N V 1239 & 121 & 138 & 12.3 & 138 & 90.5 & 52.5  \\
		N V 1243 & 60.2 &  67.1 & 10.3 & 69.2 & 50.2 & 37.8  \\
		Fe XII 1242 & 8.78 &  5.92 & 48.3 & ...  & ... & ...  \\
		H I (Ly$\alpha$) 1215.6 & 1.27E5 &  5.88E4\tablenotemark{\footnotesize{$\mathrm{c}$}} & 116 & 8.89E4 & 3.18E4\tablenotemark{\footnotesize{$\mathrm{c}$}} & 180  \\ 
		\enddata
		\tablenotetext{\footnotesize{\textnormal{a}}}{ Error is defined as percentage fractional difference, $\displaystyle\frac{|I_{\text{observed}} - I_{\text{synthetic}}|}{I_{\text{observed}}}\cdot100\%$}. This fractional difference exceeds measurement uncertainty in all cases except Si IV and C IV in GJ 581 spectra, where they are roughly equal \citep{lit:Youngblood2016}
		\vspace*{-5pt}
		\tablenotetext{\footnotesize{\textnormal{b}}}{ Observed fluxes were increased by 30\% to account for interstellar absorption}
		\vspace*{-5pt}
		\tablenotetext{\footnotesize{\textnormal{c}}}{ Reconstructed Ly$\alpha$ flux from the MUSCLES portal}
		\vspace*{-20pt}
		
	\end{deluxetable*}
	
	The synthetic visible spectra are compared with observations in Figure \ref{fig:opt}. Hereafter, the plots comparing synthetic and observed spectra show synthetic spectra that have been smoothed to match instrumental resolution, unless specified otherwise. While there is no discernable continuum due to an extremely high density of molecular lines formed in the photosphere, we note that the intensity of the pseudo-continuum is set by the temperature at $\tau=2/3$. This value corresponds to the source of stellar emission because the stars are unresolved, and we observe their emergent flux at the mean angle of $\mu=2/3$, where $\mu$ is defined as in the radiative transfer equation. Pseudo-continua for GJ 832 and GJ 581 form at ($P=2.7\times 10^5$ dyne cm$^{-2},$ $T=3550$ K) and ($P=4.5\times 10^5$ dyne cm$^{-2},$ $T=3450$ K), respectively. The overall shapes and intensities of the pseudo-continua are well-reproduced by our models, but we note that there is likely a missing source of molecular opacity between 4000 and 4500 \AA, leading to higher intensities in this wavelength range for both modeled stars, but especially for GJ 581. This can be attributed to lower temperatures in the upper photosphere of GJ 581 compared to GJ 832, leading to higher populations of molecular species.
	
	The synthesized chromospheric lines in the visible spectra generally match the observations well, particularly the Na I D doublet. In our models, this line forms at the base of the first chromospheric rise and is mostly affected by pressure and temperature at this layer of the atmosphere. This line was also well-matched in \citet{lit:Fontenla832}, but due to the faulty calibration procedure described above, the continuum was too bright and the observed line profile was weaker in absorption and, accordingly, the temperature minimum occurred at a higher pressure than in the present model. The first chromospheric rise in the atmosphere of GJ 581 occurs at an even lower pressure (Figure \ref{fig:models}) and lower temperature, hence the deeper absorption profile. This is consistent with previous work by \citet{lit:2000Mauas}.

	GJ 832 and GJ 581 have H$\alpha$ in absorption, indicating that both stars are not very active as seen at optical wavelengths. We find good agreement between synthetic and observed H$\alpha$ profiles in GJ 832 spectra (Figure \ref{fig:opt}). In GJ 581, however, the observed H$\alpha$ profile is weak in absorption compared to GJ 832, while the synthetic profile shows neither emission nor absorption, possibly because it is filled in by overlapping molecular opacities. The relatively low ($\sim$-5.7 for GJ 581 and $\sim$-5.1 for GJ 832) values of the activity indicator $\log R'_{HK}$, defined as the ratio between Ca II H \& K chromospheric emission and bolometric flux, also suggest that the stars are not very active \citep{lit:2020Melbourne,lit:2017AsDe}.
	
	Ca II H \& K lines fit reasonably well in the case of GJ 832. Following the procedure in \citet{lit:Fontenla832}, we merged the levels of Ca II and Mg II ions that have the same quantum numbers, except the total angular momentum, and assumed that the merged levels are in LTE with respect to one another. This method is useful for decreasing computing time, and only affects the relative fluxes of H and K lines, hence the discrepancy between observed and synthesized Ca II profiles. The combined fluxes of the two lines are similar between the synthetic and the observed spectra (see Table \ref{tab:lines}). In the case of GJ 581, the CASLEO spectrum does not have a high enough signal-to-noise ratio to discern the profiles of the Ca II H \& K lines. We therefore did not use Ca II doublet as a constraint for the model of GJ 581. However, we report computed fluxes for these lines in Table \ref{tab:lines}. We note that the ratio between computed total Ca II fluxes in GJ 832 and GJ 581 is similar to the ratio between equivalent widths of Ca II K lines for these stars reported in \cite{lit:Youngblood2017}.
	
	We compute contribution functions for several key lines, including the Ca II doublet. The contribution function at a frequency $\nu$ is defined as $\displaystyle f_{\nu} = \frac{c^2}{2h\nu}\varepsilon \text{e}^{-\tau_{\nu}}$, where $\varepsilon$ is the total line emissivity and $h$ is the Planck constant. We use the dependence of the contribution function on temperature to make precise changes in narrow layers of the atmospheric profiles. In Figure \ref{fig:contrib_opt}, we show the contribution functions of the Ca II lines. Both the emission peaks (H$_2$ and K$_2$ features -- not to be confused with molecules) and the self-reversed cores (H$_3$ and K$_3$) are formed in the upper chromosphere and lower TR, thereby providing a useful diagnostic for this region. We note that the contribution function of Ca II for GJ 581 is lower compared to GJ 832, which explains the relative strength of Ca II doublet in the spectra of these stars.
	
	\subsection{Near Ultraviolet Spectrum} \label{sec:spectra_nuv}
	
	\begin{figure}
	\gridline{\fig{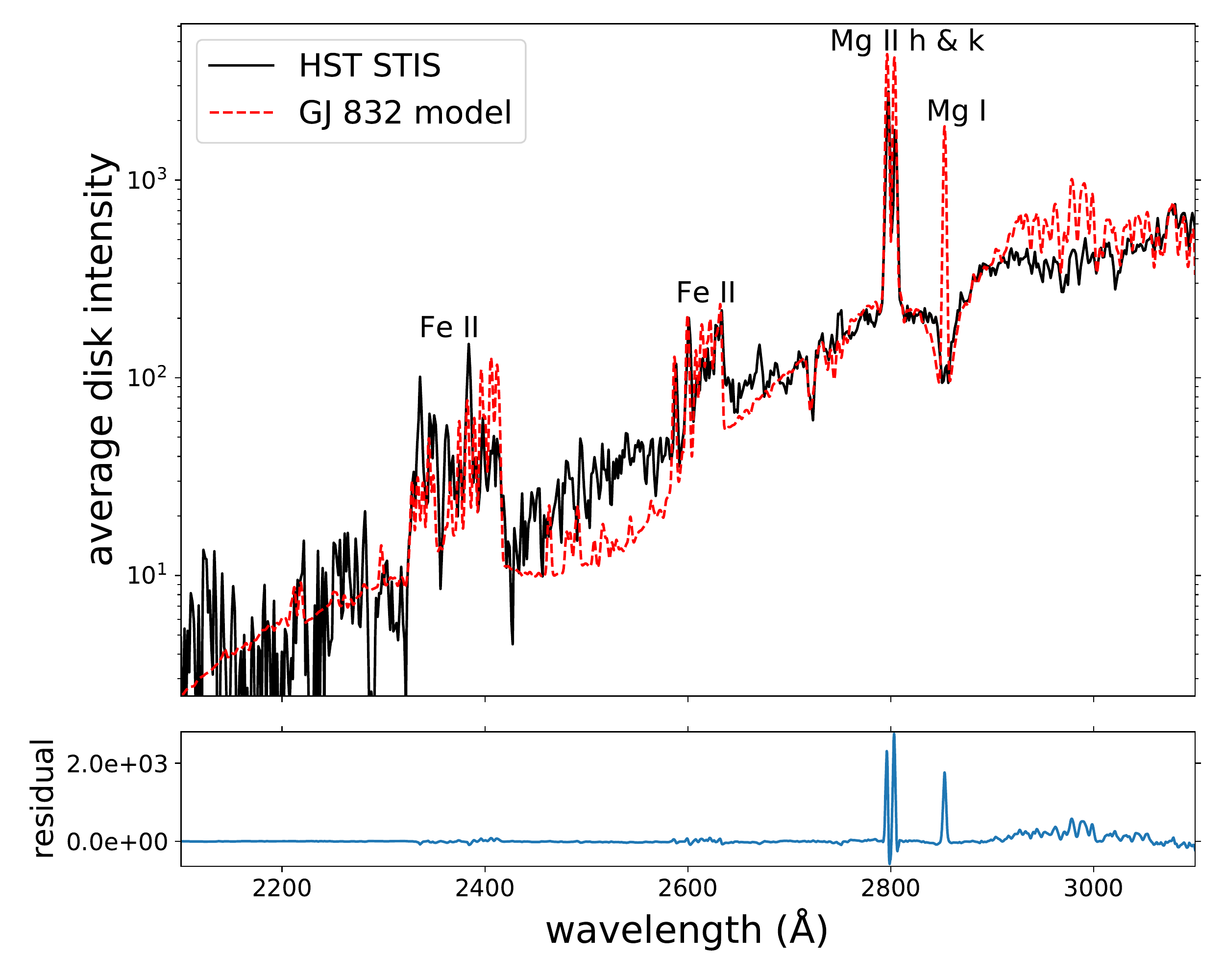}{0.5\textwidth}{}
		\vspace{-30pt}
		}
	\gridline{\fig{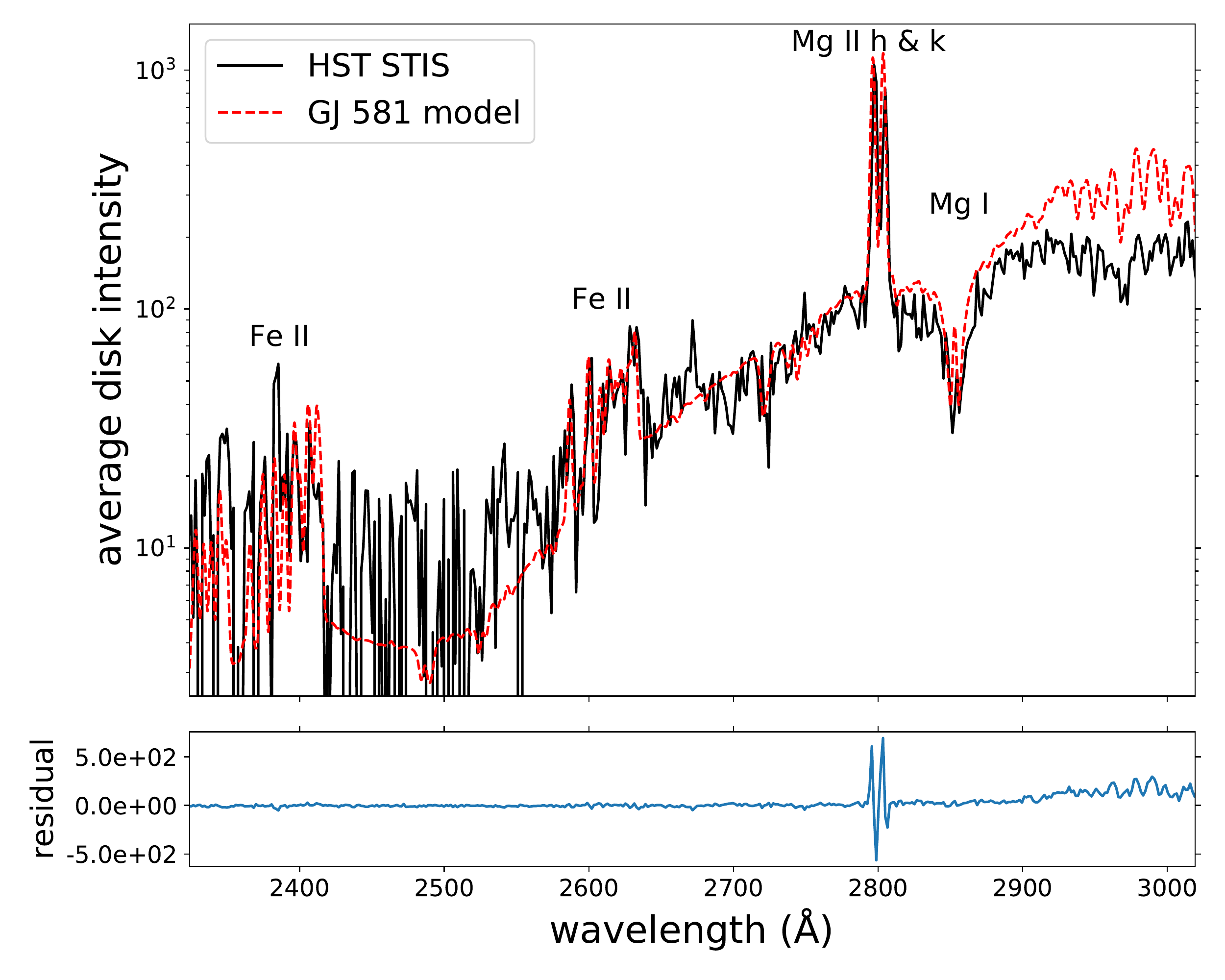}{0.5\textwidth}{}
		\vspace{-20pt}
		}
	\caption{Comparison between observed (black) and computed (red) average disk intensities (erg cm$^{-2}$\AA$^{-1}$s$^{-1}$sr$^{-1}$) of GJ 832 (top) and GJ 581 (bottom) in the near-UV range.
		\label{fig:nuv}}
	\end{figure}

	\begin{figure}
	\gridline{\fig{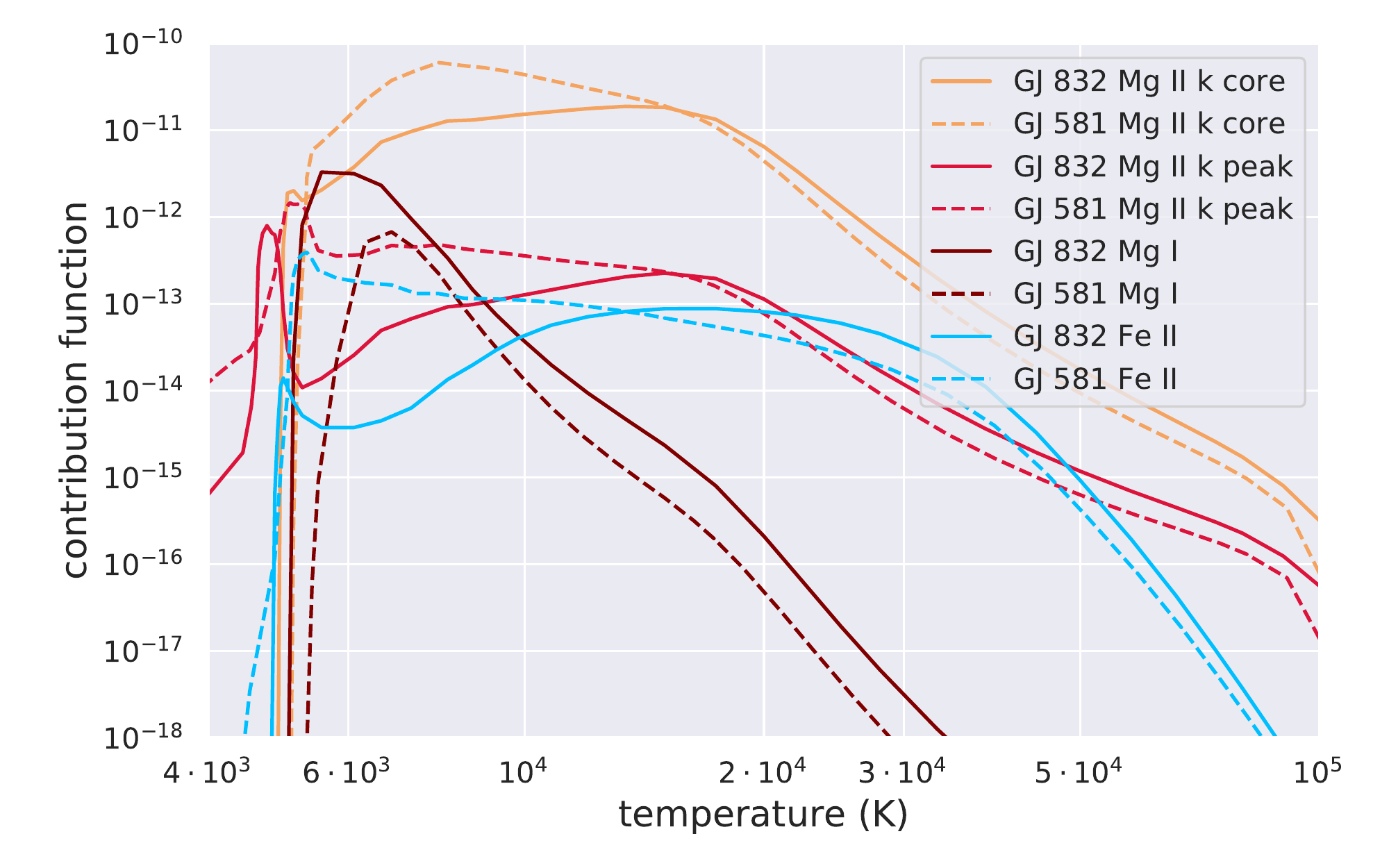}{0.5\textwidth}{}
		\vspace*{-20pt}
	}
	\caption{Contribution functions of NUV lines. Solid (dashed) lines represent contribution functions in the model of GJ 832 (GJ 581).
		\label{fig:contrib_nuv}}
	\end{figure}
	
	We used near-UV spectral data from the MUSCLES Treasury Survey \citep{lit:Loyd2016}, specifically, the spectra obtained with the G230L grating on the \textit{HST (Hubble Space Telescope)} STIS (Space Telescope Imaging Spectrograph) instrument. All MUSCLES data products used in the present paper are version 2.2. For the computation of NUV spectra, we assumed plane-parallel geometry and used the same stellar radii as we did for the visible spectra. We also applied appropriate Doppler shift corrections of ~$\approx 0.5$ \AA\ to the STIS G230L spectra of GJ 581 to match the Mg II h \& k profiles. The comparison between synthetic and observed NUV spectra is shown in Figure \ref{fig:nuv}.
	
\begin{figure}
		\gridline{\fig{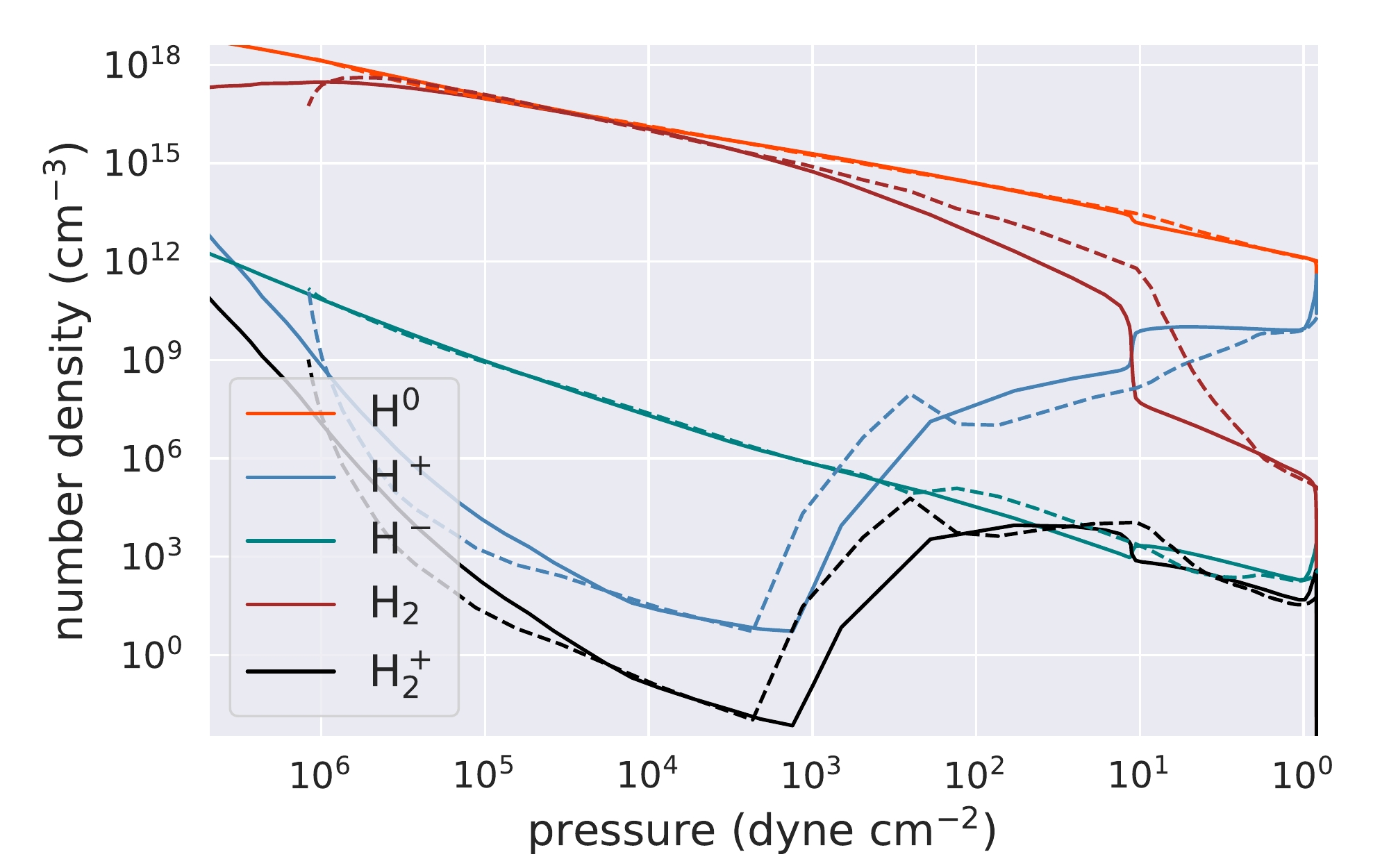}{0.5\textwidth}{}
			\vspace*{-20pt}
		}
		\caption{Computed number densities of hydrogen species in the atmospheres of GJ 832 (solid lines) and GJ 581 (dashed lines). Hydrogen molecules are computed assuming chemical equilibrium.
			\label{fig:hydrogen}}
	\end{figure}

	We find a very good match between computed and observed NUV spectra for GJ 581 and a reasonably good match for GJ 832. The NUV continua of GJ 832 and GJ 581 form at ($P\approx700$ dyne cm$^{-2},$ $T\approx2655$ K) and ($P\approx500$ dyne cm$^{-2},$ $T\approx2550$ K), respectively. To get correct NUV intensities, we find it critical to include opacities of molecular species, such as CH, NH, OH, and SiH. Another important source of opacity is H$_2$, which in our models is calculated simultaneously with H$^-$, H$_2^+$, H$^+$, and H$^0$ assuming chemical equilibrium. This assumption, however, may not hold near the temperature minimum \citep{lit:SRPM2015}. We find that the opacity of H$_2$ affects the intensity of NUV continuum. In particular, we used a higher \textit{ad hoc} H$_2$ opacity for GJ 581 than for GJ 832, while leaving H$_2$ densities unmodified. We justify this by noting that the density of H$_2$ in the NUV formation region of GJ 581 exceeds that of GJ 832 (Figure \ref{fig:hydrogen}). Other than the continuum, NUV spectra do not respond to changes in this parameter.
	
	The Fe II multiplets are also in excellent agreement with observations. As shown in Figure \ref{fig:contrib_nuv}, these lines form in the upper chromosphere and, along with Mg II doublet, can facilitate photolysis of HO$_2$, H$_2$O$_2$, and O$_3$ in exoplanetary atmospheres \citep{lit:2014Tian}.
	
	The Mg II doublet is one of the brightest UV features in M dwarf spectra. These are optically thick lines that form in the chromosphere and TR over a wide range of temperatures: the line cores form around 10,000 K, while the wings form at lower temperatures in the chromospheric plateau. It is, therefore, a spectral feature that can be used to constrain the thermal structure of the chromosphere and lower TR. \citet{lit:2013France} used high-resolution STIS E230H spectra of GJ 832 to show that the interstellar absorption removes 30-35\% of the intrinsic Mg II flux. GJ 581 was not observed in this regime, but the H I column density of the ISM clouds in its line of sight (LOS) is intermediate between GJ 832 and GJ 667C \citep[see Table 2 in ][]{lit:2013France}, both of which are predicted to have the Mg II lines attenuated by 30-35\%. Further, the magnitudes of radial velocities of GJ 832 and GJ 581 compared to their respective LOS ISM clouds are close to one another: 18 km/s and 24 km/s, respectively \citep{lit:ISM,lit:gaia}. Therefore, it is reasonable to assume that the Mg II doublet in GJ 581 spectra is also attenuated by 30-35\%. Correcting for this interstellar absorption, our models reproduce the observed Mg II fluxes well. The Mg II lines show a strong central reversal in high-resolution computed spectra of both stars. This is because the peaks (h$_2$ and k$_2$) form at significantly lower temperatures compared to the cores (h$_3$ and k$_3$, Figure \ref{fig:contrib_nuv}), where the line source function is decreasing with height. The central reversal is degenerate with ISM absorption profile reported in \citet{lit:2013France}, meaning that said absorption for the Mg II lines can be negligible.
	
	Mg I 2582~\AA\ is a problematic line, especially in our model of GJ 832. This problem was also present in the GJ 832 model built by \citet{lit:Fontenla832}. Mg I is an absorption line in both observed stellar spectra with a weak central emission, but, for GJ 832, the line core emission is too strong. As shown in Figure \ref{fig:contrib_nuv}, Mg I is a line whose contribution functions differ between GJ 832 to GJ 581. In the GJ 832 model, the line forms at significantly higher pressures compared to GJ 581, which may explain why the fit is worse for GJ 832. The discrepancy between computed and observed Mg I line fluxes could be caused by inaccurate ionization balance or by an overestimate of magnesium abundance in the stellar atmospheres.
	
	\subsection{Far Ultraviolet Spectrum} \label{sec:spectra_fuv}
	
	\begin{figure}
		\gridline{\fig{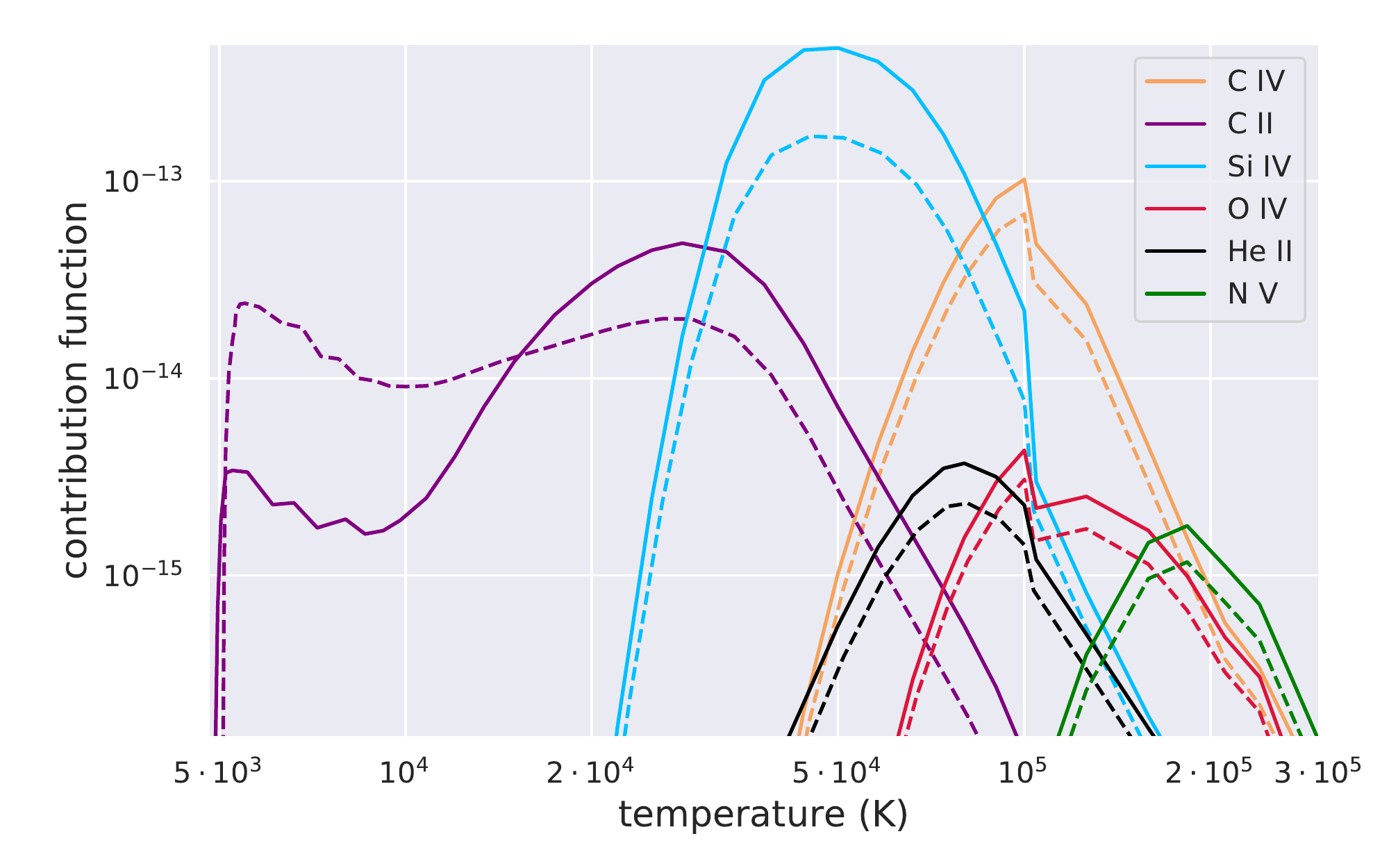}{0.5\textwidth}{}
			\vspace*{-20pt}
		}
		\caption{Contribution functions of FUV lines. Solid (dashed) lines represent contribution functions in the model of GJ 832 (GJ 581).
			\label{fig:contrib_fuv}}
	\end{figure}
	
	\begin{figure*}
		\gridline{\fig{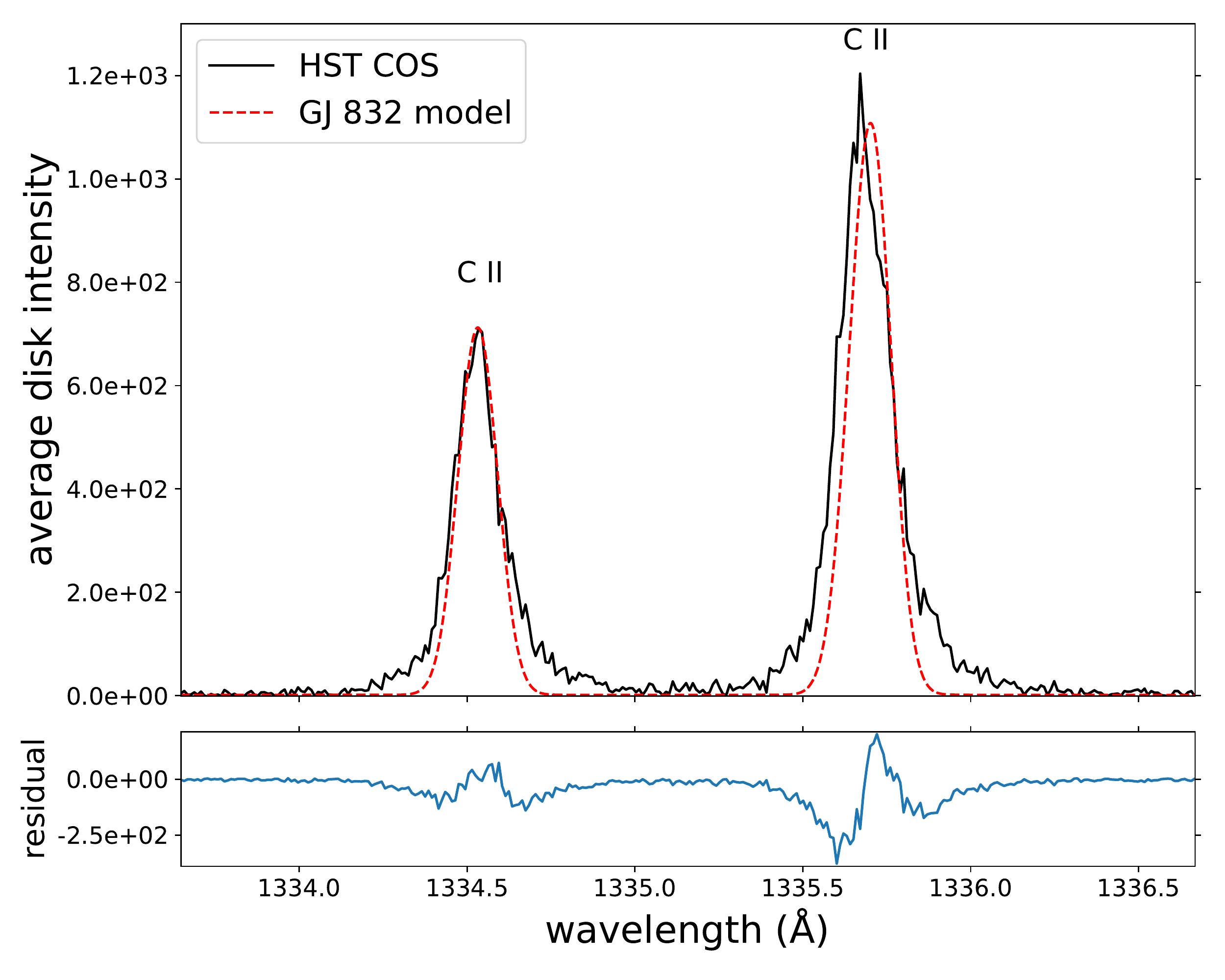}{0.45\textwidth}{}
			\fig{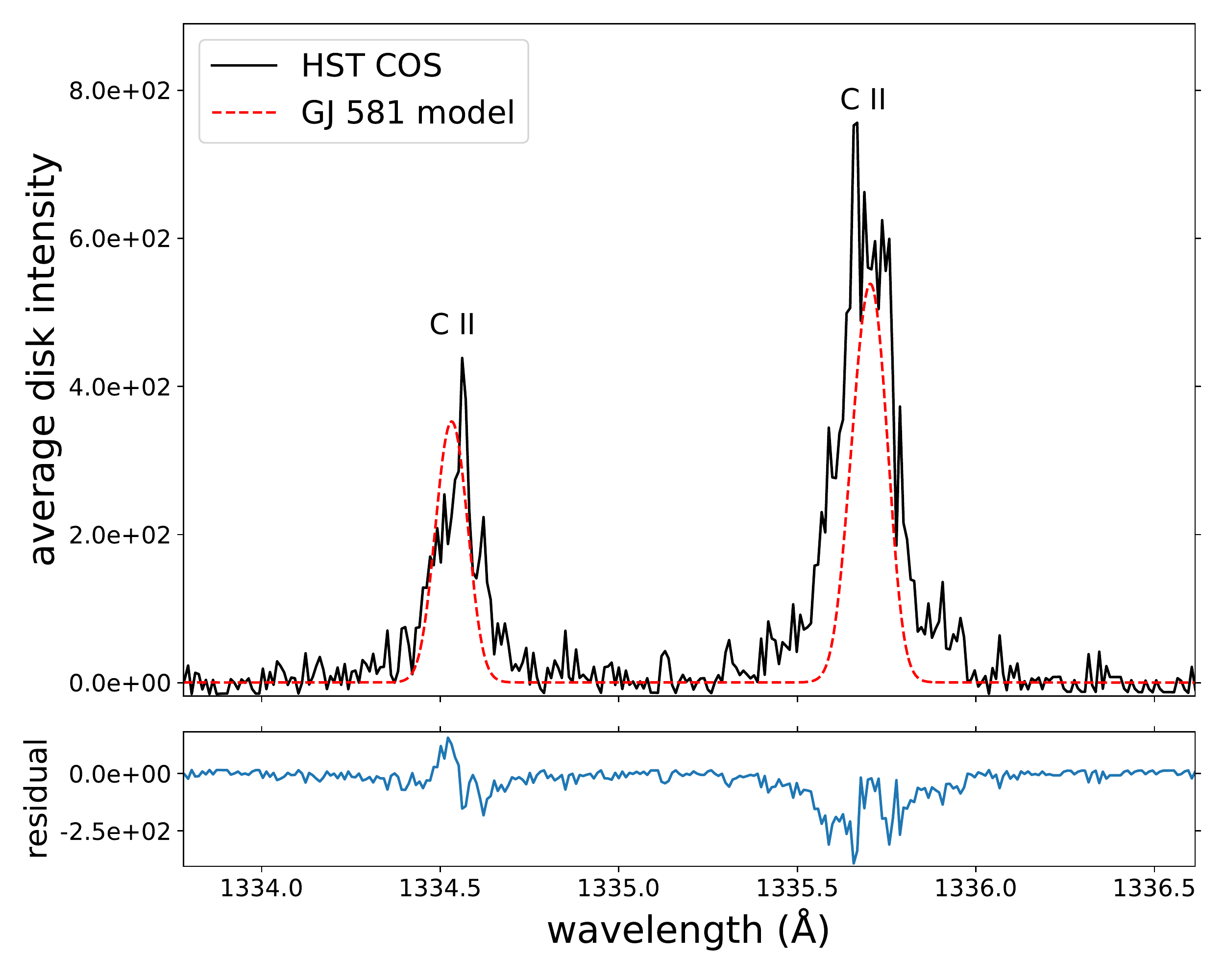}{0.45\textwidth}{}
			\vspace*{-20pt}
		}
		\gridline{\fig{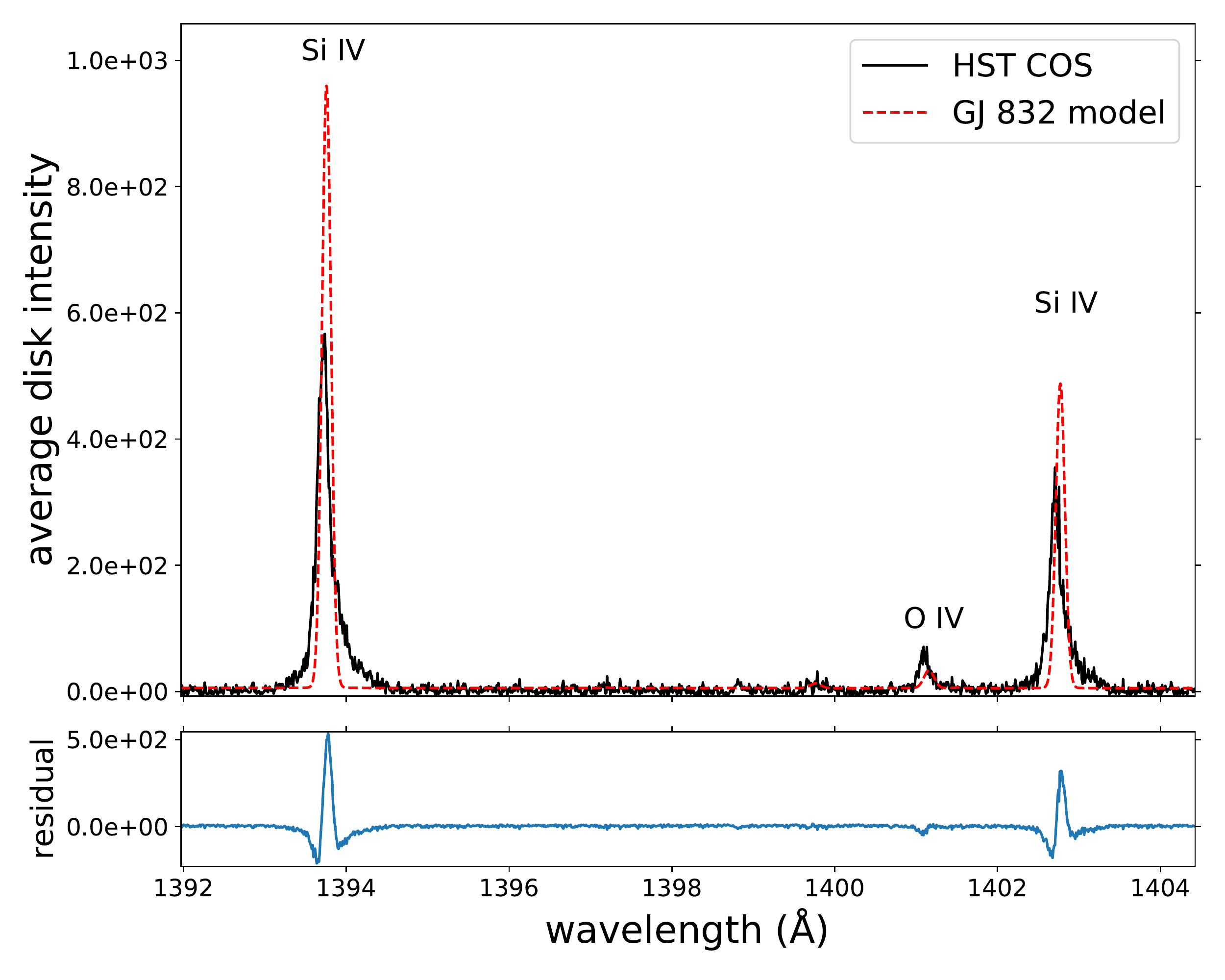}{0.45\textwidth}{}
			\fig{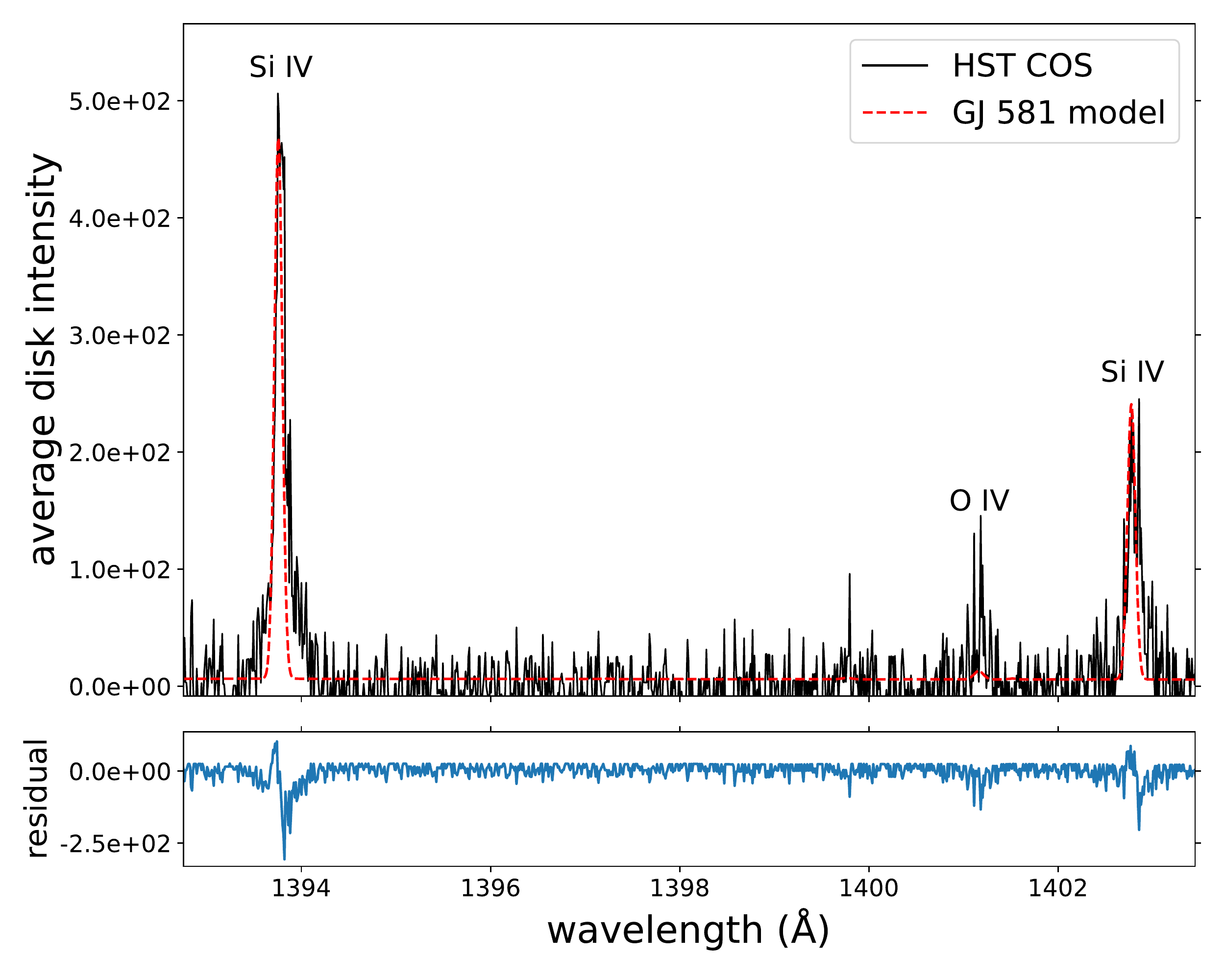}{0.45\textwidth}{}
			\vspace*{-20pt}
		}
		\gridline{\fig{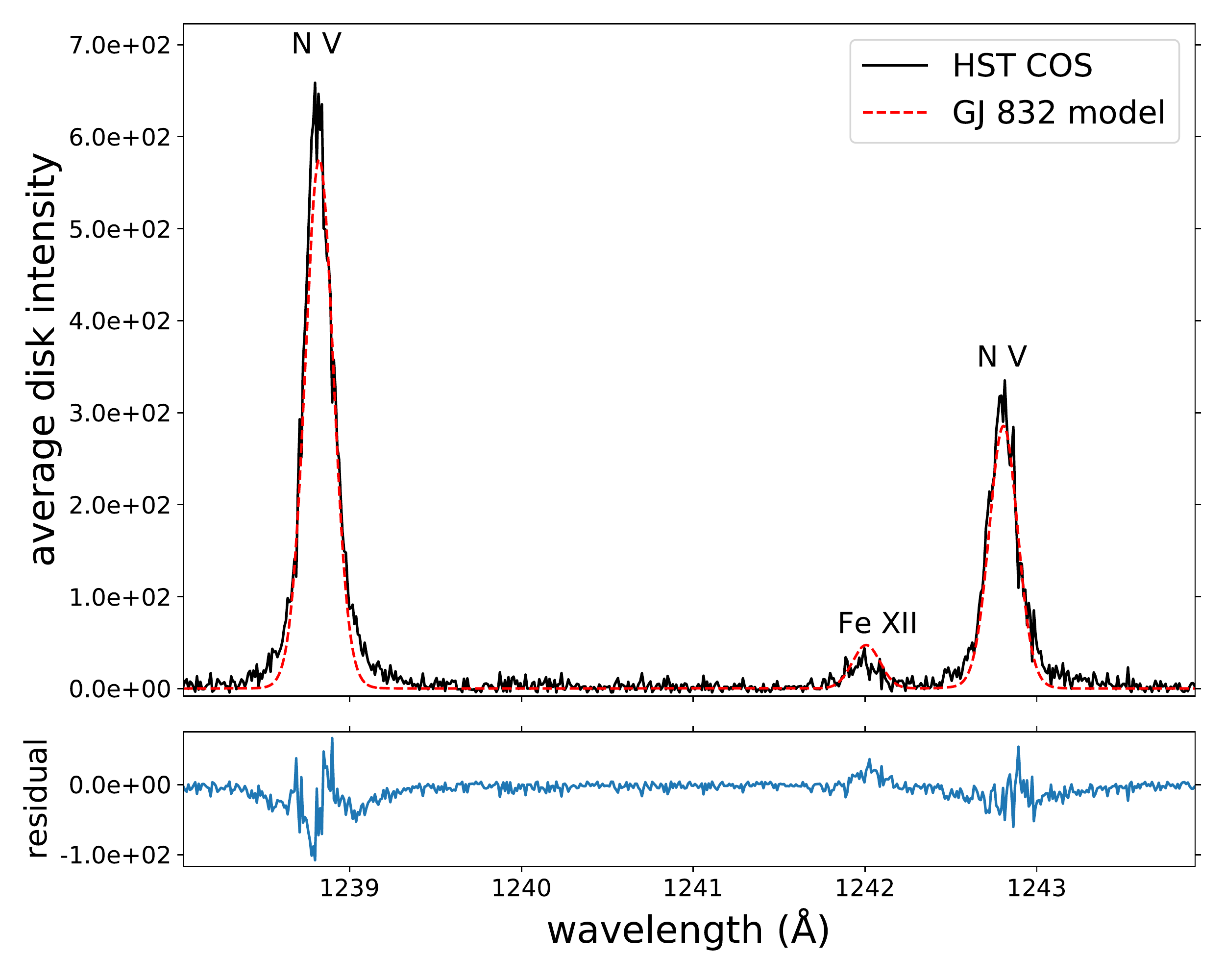}{0.45\textwidth}{}
			\fig{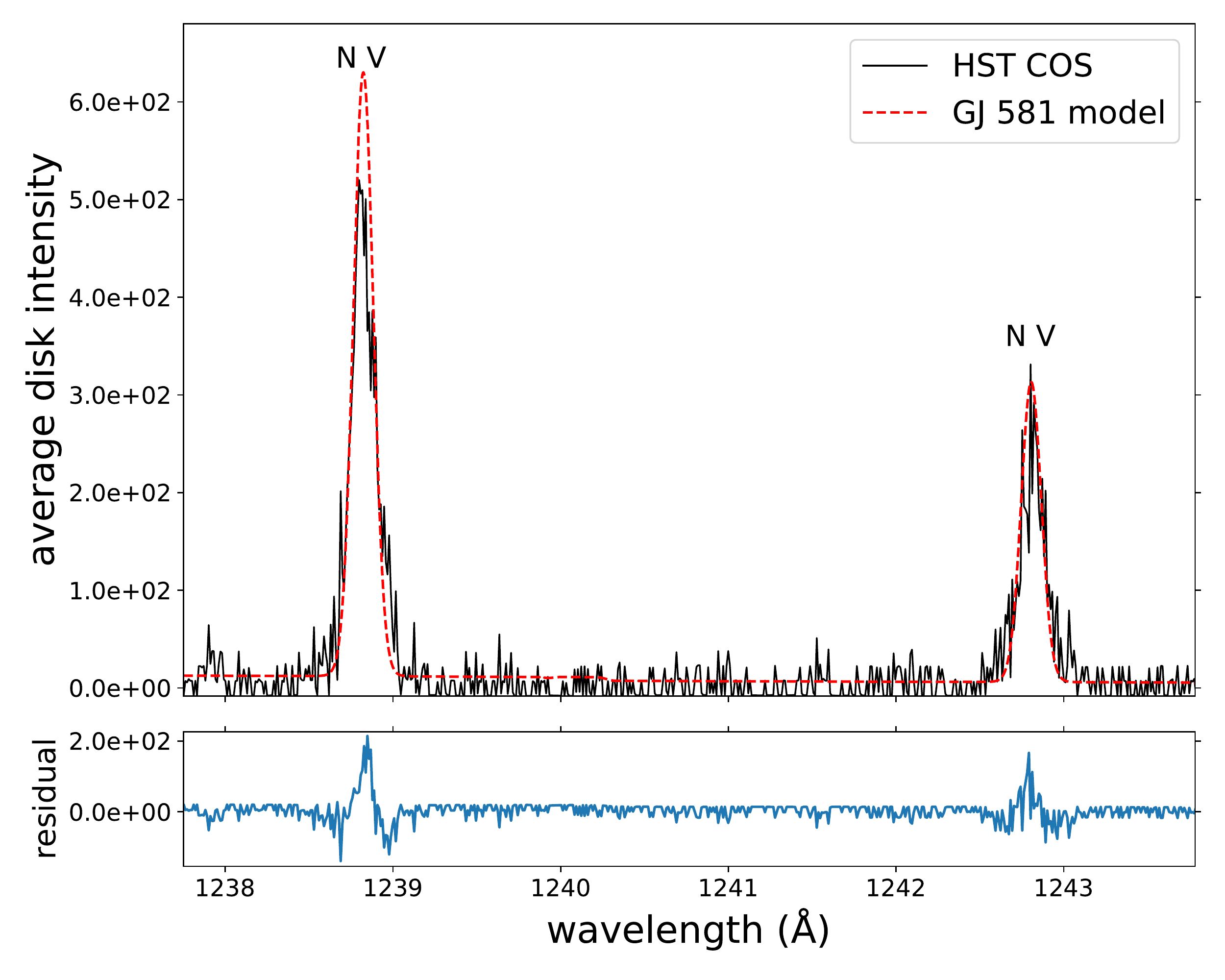}{0.45\textwidth}{}
			\vspace*{-20pt}}
		\caption{Comparison between observed (black) and computed (red) average disk intensities (erg cm$^{-2}$\AA$^{-1}$s$^{-1}$sr$^{-1}$) of GJ 832 (left column) and GJ 581 (right column) in the far-UV range. \textit{Top row:} C II doublet (1333 \AA) without accounting for ISM absorption; \textit{middle row:} Si IV doublet (1400 \AA); \textit{bottom row:} N V doublet (1240 \AA). The bottom subplots in each panel show the difference between the synthetic and observed spectra for each wavelength.
			\label{fig:fuv}}
	\end{figure*}
	
	We use MUSCLES Treasury Survey spectra that were obtained with the G130M and G160M gratings of the \textit{HST} COS (Cosmic Origins Spectrograph) instrument \citep{lit:Loyd2016}. The combined coverage of G130M and G160M spans the range between 1150 and 1775 \AA\ with spectral resolution between $R=12000$ and $20000$. We compare the synthesized Ly$\alpha$ profiles with the reconstructed STIS spectra from \citet{lit:Youngblood2016}.
	
	The C II and Si IV doublets provide strong constraints on the lower-to-mid TR, particularly its temperature gradient and pressure, as they are formed around 30,000 K and 60,000 K, respectively (Figure \ref{fig:contrib_fuv}). The relative weakness of the C II and Si IV lines in GJ 581 compared to GJ 832 can be attributed to lower TR pressures in GJ 581. We note that the C II 1334 line is subject to interstellar absorption and is therefore attenuated by $\sim$30\%. Additionally, flares were not removed from GJ 832 co-added spectra observed with G130M. Doppler shifts in the COS spectra were corrected for: the majority of lines are red-shifted by 0.05--0.1 \AA, with the exception of Si IV lines in GJ 832 spectra that are blue-shifted by 0.06 \AA. 
	
	Several bright FUV lines serve as constraints on the upper TR and the corona; these lines include C IV (formed around 100,000 K), O IV (160,000 K), and N V (180,000 K), which are all in good agreement with observed spectra (see Figure \ref{fig:fuv} and Table \ref{tab:lines}). Fe XII at 1242 \AA\ is detected in the spectra of GJ 832 and provides a useful diagnostic for the lower corona, since it is formed at around 1.5 MK.
	
	\begin{figure}
		\gridline{\fig{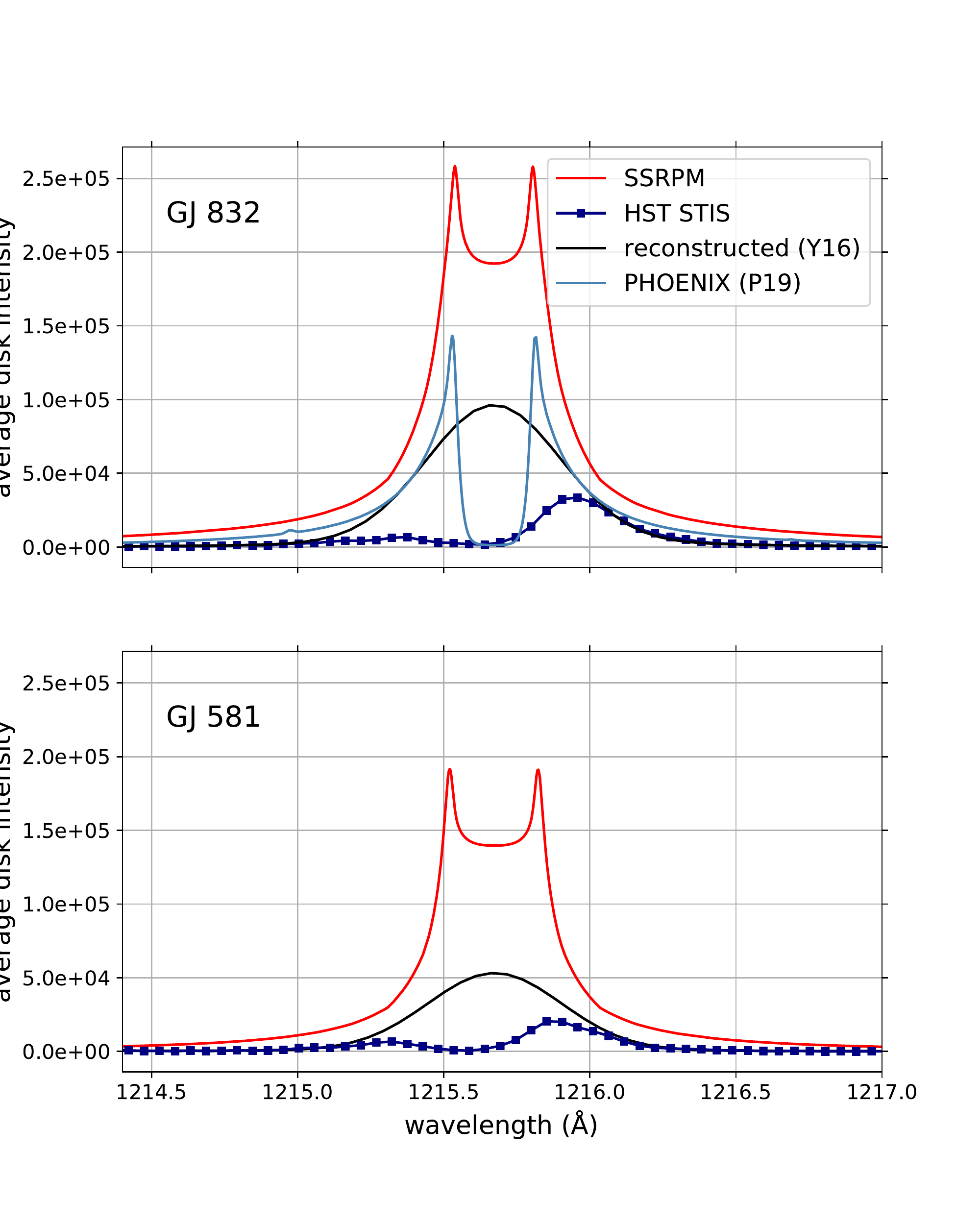}{0.5\textwidth}{}
			\vspace*{-30pt}
		}
		\caption{Comparison between computed Ly$\alpha$ profiles and reconstructed profiles from \cite{lit:Youngblood2016} (denoted as Y16). Also included is the Ly$\alpha$ profile from \citet{lit:Peacock832} (P19) for GJ 832. STIS G140M spectra, not corrected for ISM absorption, show that our models overestimate emission from Ly$\alpha$ wings.
			\label{fig:lya}}
	\end{figure}
	
	The optically thick resonance line Ly$\alpha$ is the brightest UV feature in stellar spectra. Apart from driving photochemical processes in planetary atmospheres, it also back illuminates deeper layers of stellar atmosphere, thereby enhancing H$_2$ fluorescence and potentially altering the ionization balance of heavy elements \citep{lit:2013France,lit:SRPM2015,lit:2017Kruczek}. Stellar Ly$\alpha$ lines are strongly absorbed by the ISM, hence it is important to obtain accurate profiles of this line. Our computed Ly$\alpha$ fluxes exceed those estimated by \cite{lit:Youngblood2016} (see Table \ref{tab:lines} and Figure \ref{fig:lya}) by a factor of 2--3, and the width of Ly$\alpha$ wings is likewise overestimated. In this sense, we were not able to make meaningful improvements upon the work of \citet{lit:Fontenla832}. The wide wings may be due to errors in the bound-free opacities and recombination rates of several atomic species, such as C I \citep{lit:FAL2014}, while the anomalously high fluxes likely indicate a problem with the heat balance in the TR. We also observe strong reversal cores in the Ly$\alpha$ profiles of both stars, which are not reported by \citet{lit:Youngblood2016}, although they are consistent with the previous \texttt{SSRPM} model of GJ 832 \citep{lit:Fontenla832}. These reversal cores, however, are not as strong as those computed using the \texttt{PHOENIX} models \citep{lit:Peacock,lit:Peacock832}. Observations of high radial velocity M dwarfs did not reveal such absorption cores \citep{lit:2016Guinan,lit:2019Schneider}. This points to a common underlying issue with the treatment of this line by different radiative transfer codes.
	
	\subsection{X-Ray Spectrum} \label{sec:spectra_xray}
	
	\begin{figure}
		\gridline{\fig{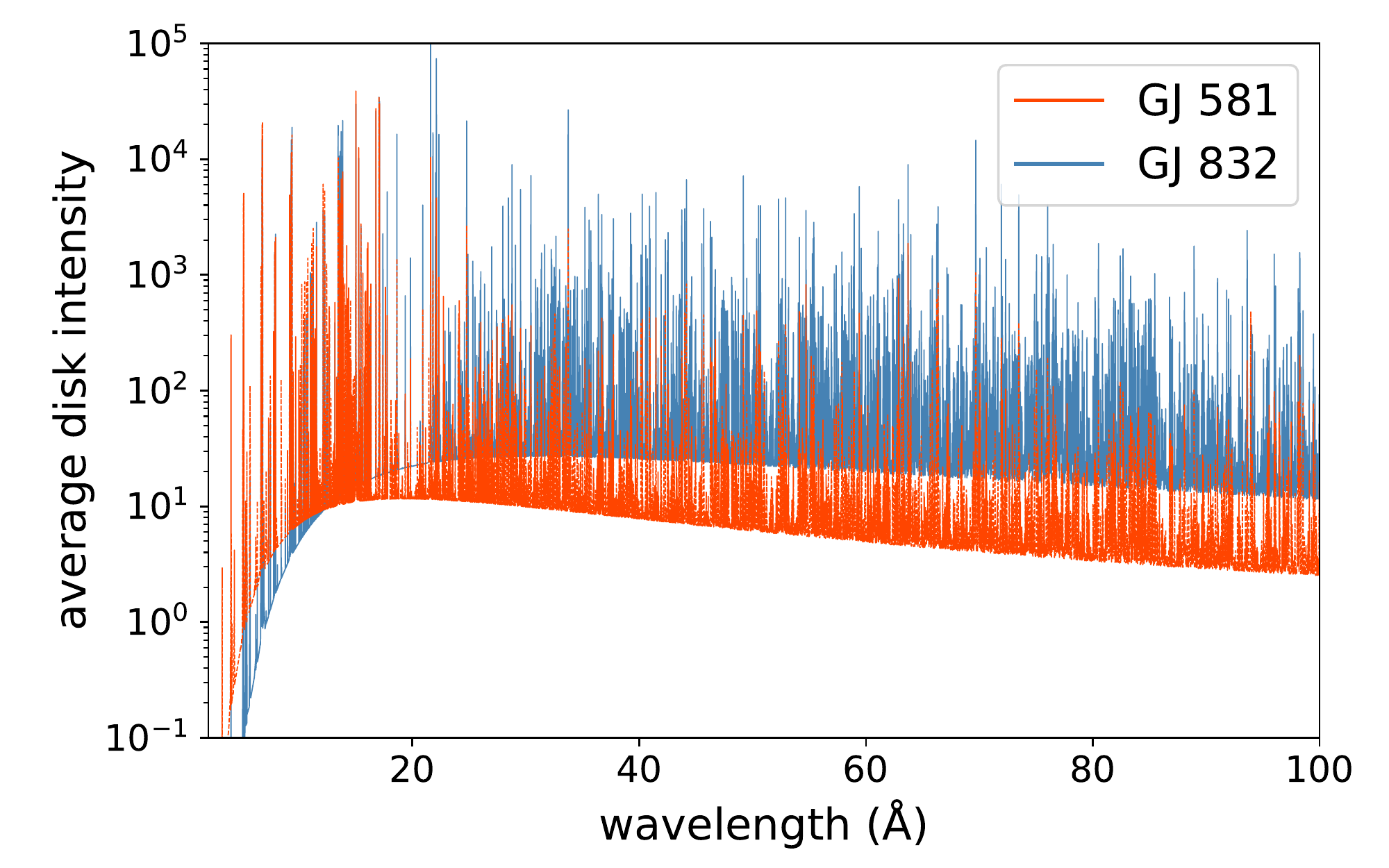}{0.5\textwidth}{}
			\vspace*{-20pt}
		}
		\caption{X-ray average disk intensities (erg cm$^{-2}$\AA$^{-1}$s$^{-1}$sr$^{-1}$) of GJ 832 (blue) and GJ 581 (red).
			\label{fig:xray}}
	\end{figure}
	
	The X-ray spectra are computed using spherical symmetry, since stellar coronae extend significantly beyond the chromospheres and are comparable in size to stellar radii. \citet{lit:Loyd2016} used \textit{XMM-Newton} data and the \texttt{APEC} (Astrophysical Plasma Emission Code) model fits to determine that the coronal emission from GJ 832 can be fit by two components, $kT = 0.09^{+0.02}_{-0.09}$ keV and $kT = 0.38^{+0.11}_{-0.07}$ keV (corresponding to approximately $1.04 \times 10^6$ K and $4.41 \times 10^6$ K, respectively), while in GJ 581, the X-ray flux is best fit by a single component with $kT=0.26 \pm 0.02$ keV ($T \approx 3.02 \times 10^6$ K).
	
	Synthesized X-ray spectra are shown in Figure \ref{fig:xray}. The disk-integrated X-ray luminosities of GJ 832 and GJ 581 are $L_{\text{X}, 832} = 5.34\cdot10^{26}$ erg s$^{-1}$ and $L_{\text{X}, 581} = 8.51\cdot10^{25}$ erg s$^{-1}$. The X-ray luminosity of GJ 832 was reconstructed by \citet{lit:2011Sanz} and \citet{lit:Loyd2016} to be $6.02\cdot10^{26}$ erg s$^{-1}$ and $2.01\cdot10^{26}$ erg s$^{-1}$, respectively, so our value is intermediate (see Table \ref{tab:XrayEUV}). GJ 581 was not considered in \citet{lit:2011Sanz}, but \citet{lit:Loyd2016} obtained a value of $9.49\cdot10^{25}$ erg s$^{-1}$, which is reasonably close to our value. However, their inferred coronal temperatures of GJ 581 are lower than in our models (see Figure \ref{fig:models}). Also, unlike the present work and \citet{lit:Loyd2016}, they do not include X-rays below 5 \AA\ in their analysis, but the contribution from this wavelength range to the overall X-ray luminosity is negligible.
	
	%\vspace{10pt}
	\section{Results and Discussion} \label{sec:disc}
	
	\begin{figure*}
		\gridline{\fig{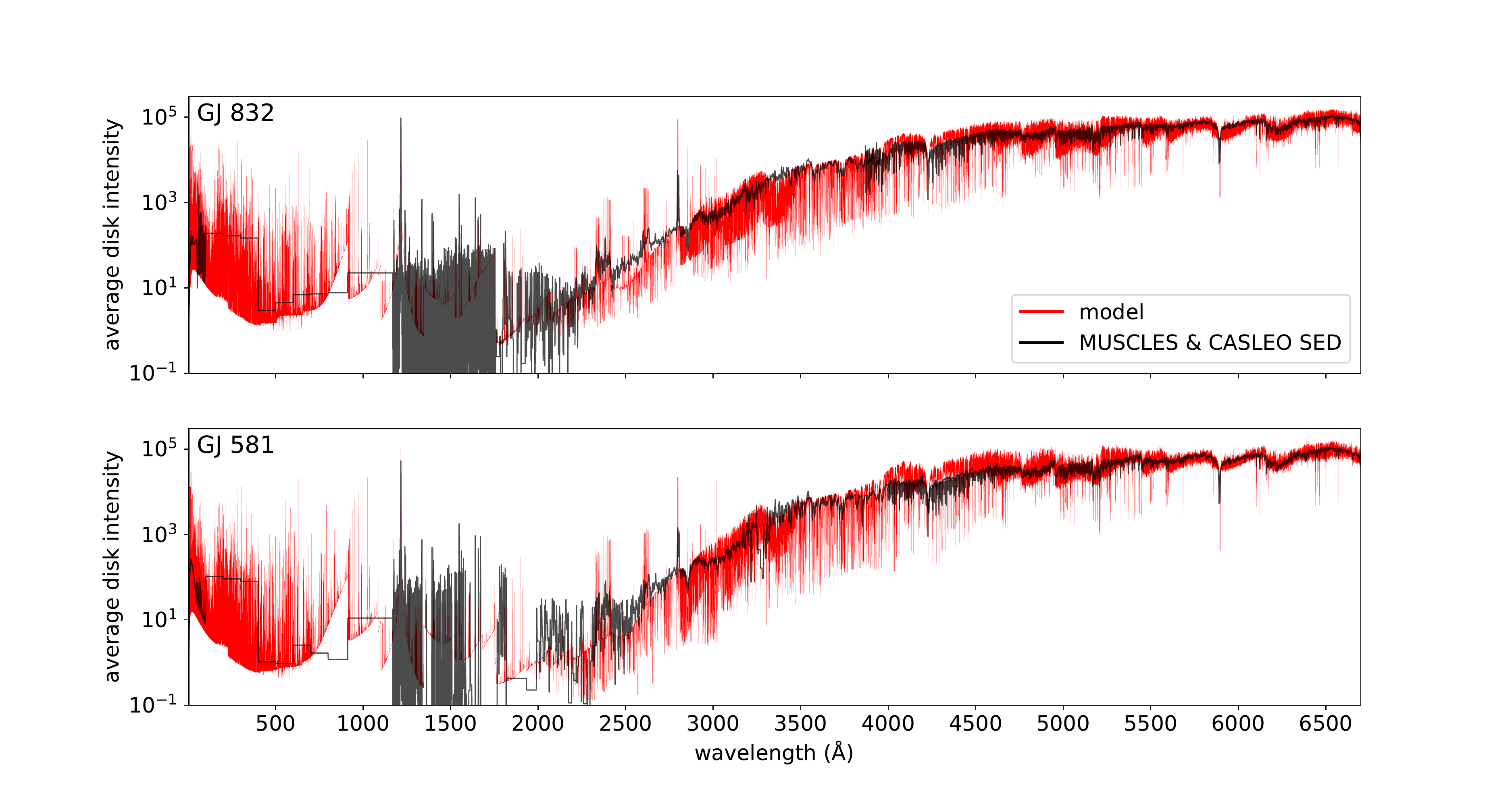}{1.05\textwidth}{}
			\vspace*{-20pt}
			%\hspace{20pt}
		}
		\caption{ X-ray--Visible synthetic spectra (red) of GJ 832 and GJ 581 compared MUSCLES and CASLEO spectra (black). The units of average disk intensity are erg cm$^{-2}$\AA$^{-1}$s$^{-1}$sr$^{-1}$. The MUSCLES spectra shown here retain instrument resolution and have been downsampled where signal-to-noise ratio is low. The synthetic spectra are also shown in \texttt{SSRPM} native resolution.
			\label{fig:panchromatic}}
	\end{figure*}
	
	We synthesized panchromatic (1 \AA\ -- 100 $\mu$m) spectra of GJ 832 and GJ 581 (Figure \ref{fig:panchromatic}). We compare their SEDs in the range 5--6700 \AA\ with CASLEO observed spectra ($\approx$3950--6700 \AA) and MUSCLES composite spectra ($\lessapprox 3950$ \AA), which were observed by several instruments aboard \textit{HST} (see Sec. \ref{sec:spectra}) or estimated using reconstruction techniques (Sec. \ref{sec:disc_euv}). The bolometric luminosities of GJ 832 and GJ 581, as well as relative fractions of fluxes in several specific bands, are shown in Table \ref{tab:luminositybyband}. The SEDs of the two stars display a certain similarity, which is not surprising, as the stars are of comparable spectral types and effective temperatures.
	
	\begin{deluxetable*}{lcccccc}
	\centerwidetable
	\tablecaption{Relative Fractions of Synthesized Stellar Luminosities in Various Bands\label{tab:luminositybyband}}
	\tablehead{
	    %\vspace{-5pt}
		\colhead{} & \colhead{Bolometric Luminosity} & \colhead{X-ray/EUV} & \colhead{FUV} & \colhead{NUV} & \colhead{Vis/NIR} & \colhead{FIR} \\
		\colhead{} & \colhead{[erg s$^{-1}$]} & \colhead{($<$91.2 nm)} & \colhead{(91.2--200 nm)} & \colhead{(200--400 nm)} & \colhead{(400 nm -- 1.6 $\mu$m)} & \colhead{(1.6--100 $\mu$m)}
	}
	\startdata
	GJ 581 (present work) & 4.12E31 & 1.03E-5 & 4.84E-5 & 2.07E-3 & 0.668 & 0.330  \\
	GJ 832 (present work) & 1.21E32 & 1.90E-5 &  7.18E-5 & 2.20E-3 & 0.623 & 0.374 \\
	GJ 832 \citep{lit:Fontenla832} & 1.38E32 & \multicolumn{2}{c}{6.87E-5$^a$} & 2.26E-3 & 0.642 & 0.356 \\
	GJ 832 \citep{lit:Peacock832}$^b$ & 1.56E32 & 4.55E-5$^c$ & 5.39E-5 & 2.54E-3 & 0.683 & 0.314
	\enddata
	\vspace{5pt}
	\tablenotetext{\footnotesize{\textnormal{a}}}{ fraction of combined flux below 200 nm}
	\vspace{-5pt}
	\tablenotetext{\footnotesize{\textnormal{b}}}{ these are the revised values -- see discussion in Section \ref{sec:disc_euv}}
	\vspace{-5pt}
	\tablenotetext{\footnotesize{\textnormal{c}}}{ in their work, the X-ray range ($<$100 \AA) is not included }
	\vspace{-20pt}
	
\end{deluxetable*}
	
	\subsection{Far-UV Continuum} \label{sec:disc_fuvcont}
	
	\begin{figure}[t]
		\gridline{\fig{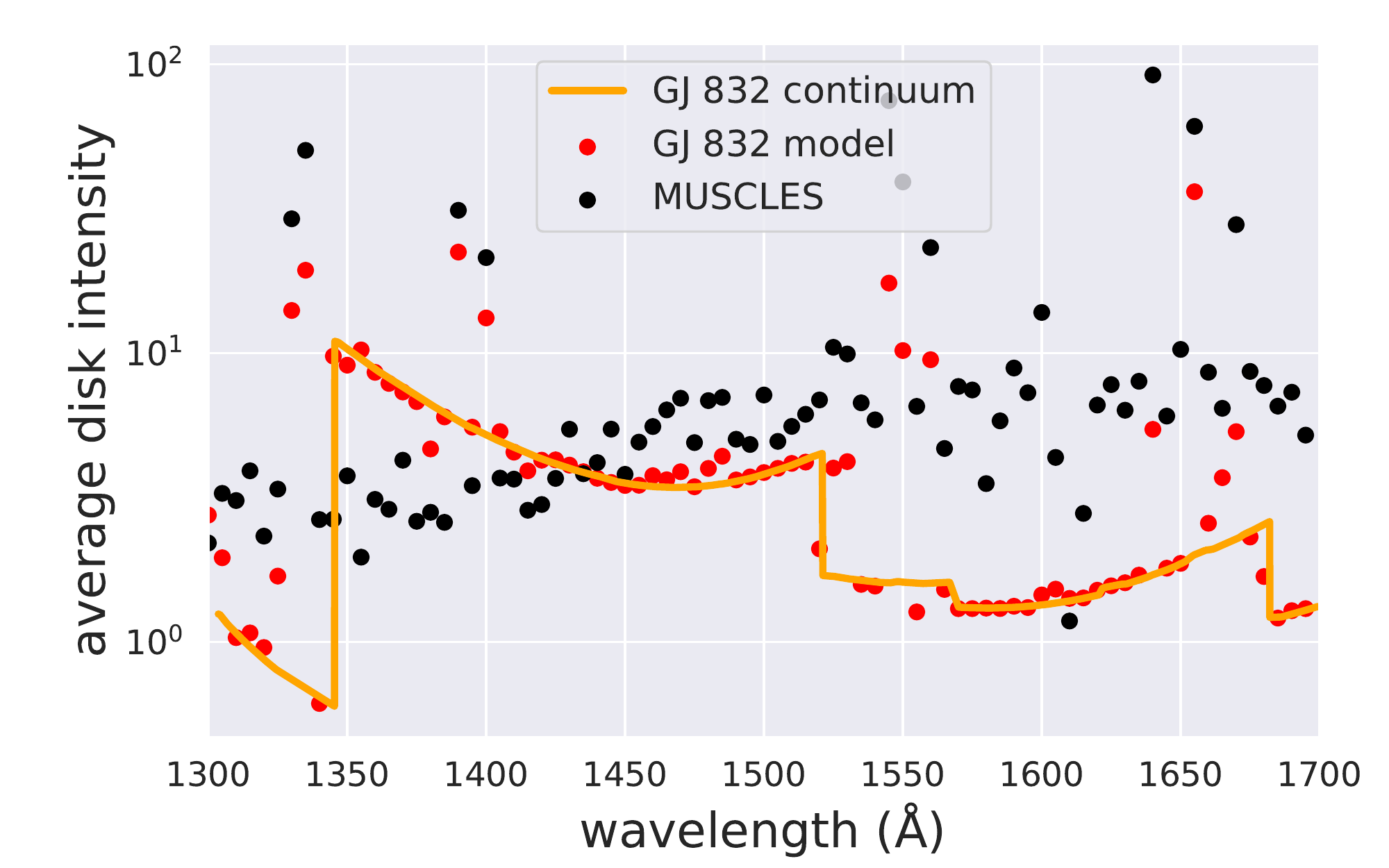}{0.5\textwidth}{}
			\vspace{-20pt}
		}
		\caption{Contribution of the FUV continuum to overall FUV emission in the range 1300--1700 \AA. Model spectra and MUSCLES spectra are binned to 5 \AA\ bins.
			\label{fig:fuv_cont}}
	\end{figure}
	
	The Far-UV continuum has only been detected in a handful of M dwarfs, and its contribution to the overall FUV emission is difficult to constrain. \citet{lit:Loyd2016} put a lower boundary of 12.2\% on the contribution of the FUV continuum for GJ 832 in the 1300--1700 \AA\ wavelength range, as they did not account for continuum flux within emission lines. They were not able to detect the FUV continuum of GJ 581 in the same range, although they did not rule out that it could also be around 10\%. Recently, \citet{lit:2020Becker} reported a photometric UV continuum detection in TRAPPIST-1 spectra using the \textit{Swift} UVOT (UV/Optical Telescope) uvw2 filter centered at 1928 \AA. However, the uvw2 filter has a FWHM bandpass of 657 \AA, so this photometric result includes emission lines and can only serve as a proxy for continuum intensity. Emission at these wavelengths can drive photolysis of water, carbon dioxide, and molecular oxygen in planetary atmospheres, hence it is important to accurately quantify it.
	
	The far-UV continua in the observed spectra of GJ 832 and GJ 581 are at the noise level (Figure \ref{fig:fuv}), but are prominent in the synthesized spectra. In Figure \ref{fig:fuv_cont}, we show MUSCLES spectra that have been binned to 5 \AA\ bins to reduce the effect of noise, similarly binned synthesized spectra, and the FUV continuum in native resolution. In synthetic spectra, the ratio of FUV continuum to FUV overall emission between 1300--1700 \AA\ is 57.2\% for GJ 832 and 43.0\% for GJ 581. These are true continua, i.e., in computing them, we do not take into account the many faint FUV lines that are blended with the continua due to instrumental broadening. 
	
	Of significant concern is the apparent bound-free opacity edge just below 1350 \AA\ associated with the first excited state of S I. It is spurious because it greatly exceeds observed fluxes (Figure \ref{fig:fuv_cont}), and because the intensity at wavelengths shorter than the edge is smaller than the intensity at longer wavelengths. One explanation for this reversal from the usual pattern is that this edge forms at heights where the temperature gradient is decreasing. The calculation of this continuum edge has been addressed in previous work \citep{lit:FAL2014}. It is also worth noting that such reversed edges can sometimes be found elsewhere in the FUV continua, e.g. the Si I edge at 1682 \AA\ in solar spectra \citep{lit:FAL2014}, although an order of magnitude drop in flux is unusual. The abnormally strong S I continuum edge may be caused by issues with photoionization balance in our atomic database, but further investigation is needed to accurately diagnose the exact cause. To exclude the possibility that the S I edge is responsible for such high continuum-to-overall emission ratio, we computed the FUV continuum flux ratios in the range between 1450--1700 \AA\, and we found that the ratios were 54.4\% for GJ 832 and 41.6\% for GJ 581, i.e. they were only changed by 1–3\%. Such high intensity of FUV continua necessitates further investigation, since it can have a significant impact on planetary atmospheres. This result appears to be consistent with the high ratios of radiative losses by the UV continuum in model chromospheres (both quiescent and active) of AU Mic \citep{lit:1996Houdebine,lit:2010aHoudebine}. However, as far as we know, such high levels of FUV continuum emission in M dwarfs have not been previously reported, and, if verified, this result has important implications for planetary photochemistry.
	
	The brightness temperature of the FUV continuum ($T_{b, \text{FUV}}$) in Sun-like stars was found to be positively correlated with rotation speed \citep{lit:Linsky_fuvcont}, and thus correlated with stellar activity. Whether this relation holds for M stars is hard to say due to their relatively low FUV flux at Earth. However, using our models, we calculated the brightness temperatures in the 1300--1700 \AA\ range of the two stars and found that $T_{b, \text{FUV}}$ is higher in the GJ 832 model by about 200 K. This suggests that a similar relation between FUV continuum brightness temperature and rotation speed may hold for M dwarfs, since the rotation periods of GJ 832 and GJ 581 are $P_{832} = 45.7\pm9.3$ d and $P_{581} = 132.5\pm6.3$ d \citep{lit:2015Suarez}. \citet{lit:Linsky_fuvcont} also reported a correlation between FUV continuum flux and Si IV line intensities, but we do not observe such correlation in our sample. The absence of this correlation among M dwarfs is consistent with previous work by \citep{lit:2018France}. Overall, it is necessary to model more stars to make robust inferences about FUV continua.
	
	\subsection{Extreme-UV Spectrum} \label{sec:disc_euv}
	
\begin{figure}
		\gridline{\fig{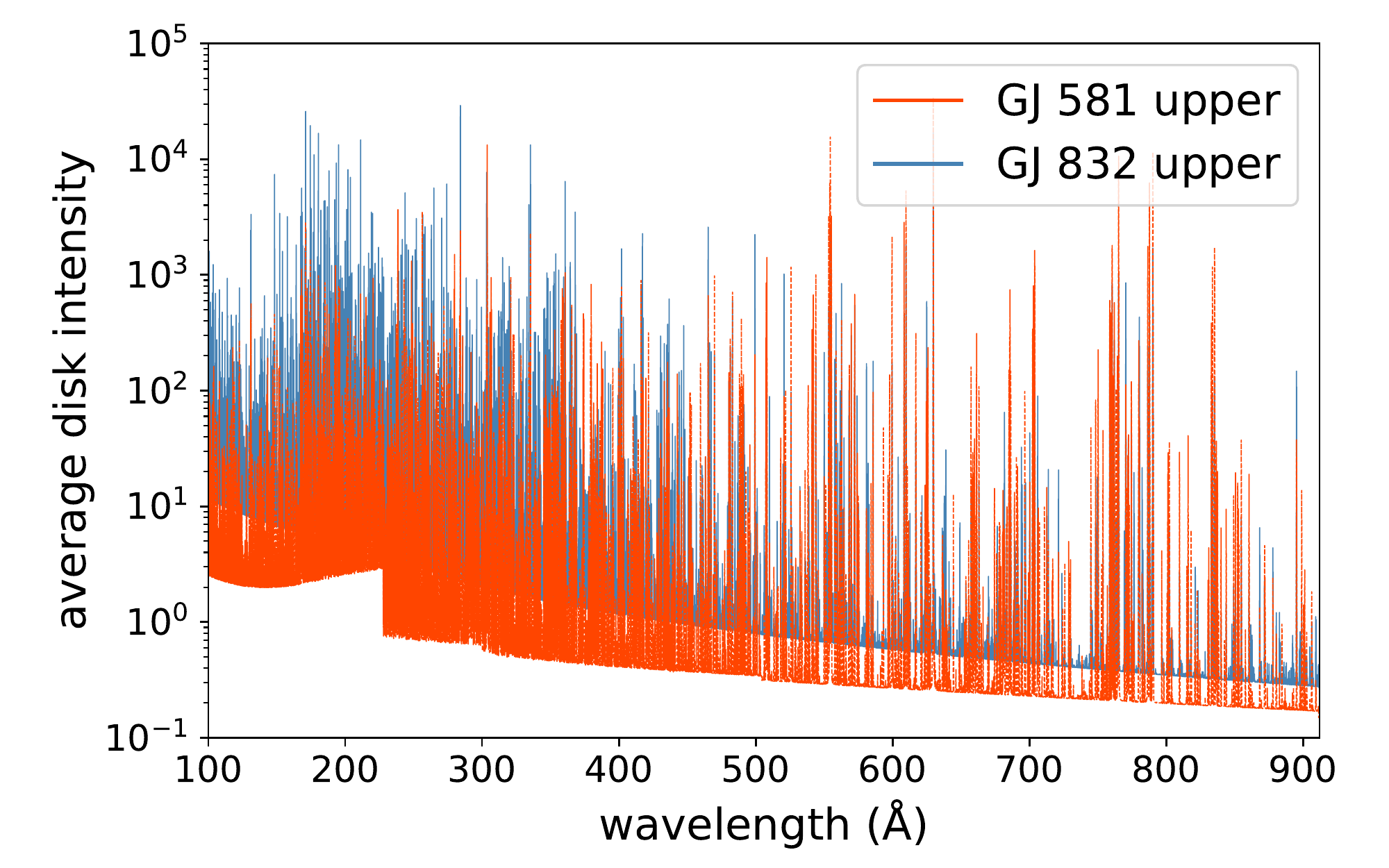}{0.5\textwidth}{}
			\vspace{-30pt}
		}
		\gridline{\fig{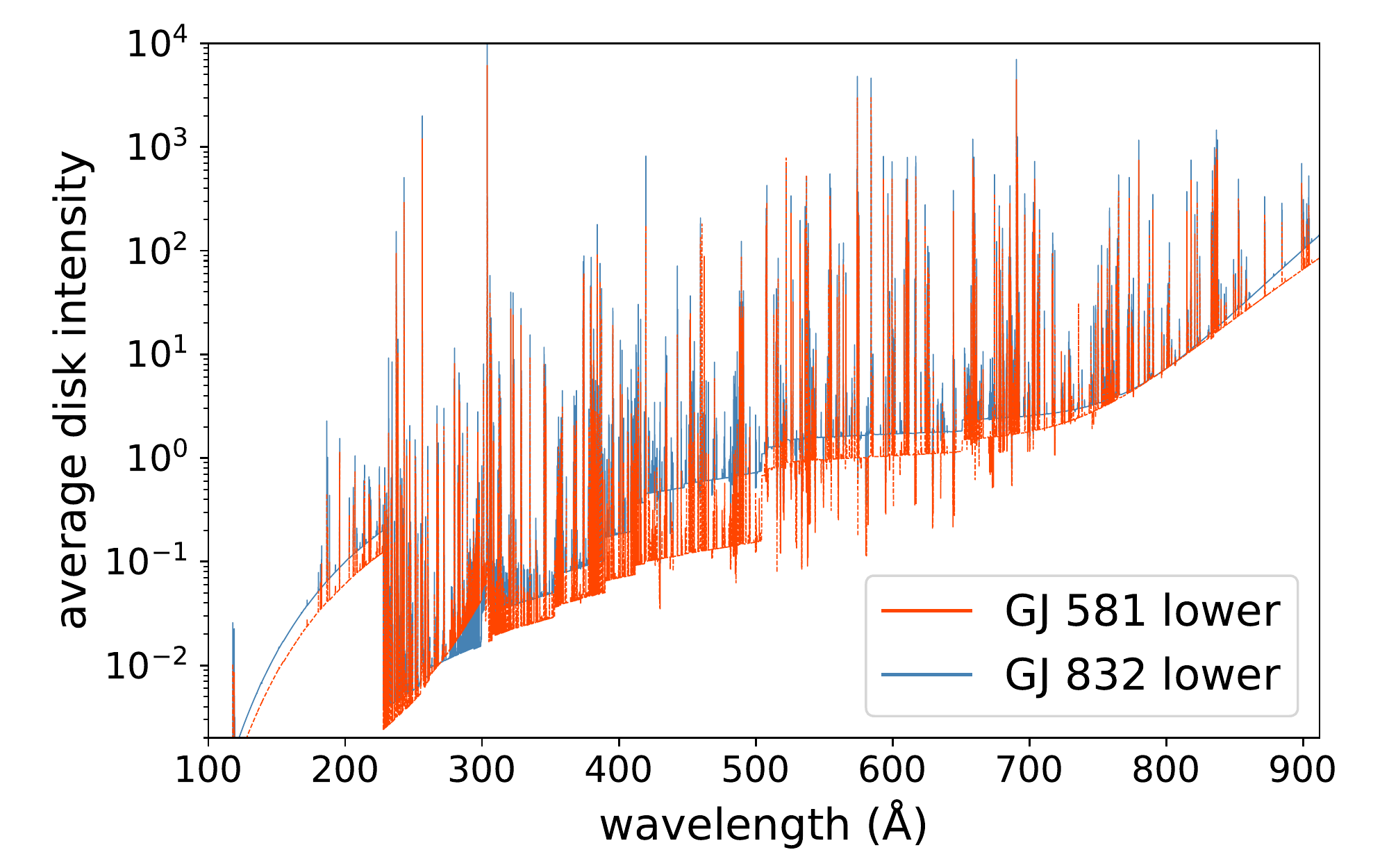}{0.50\textwidth}{}
			\vspace{-20pt}
		}
		\caption{Average disk intensity (erg cm$^{-2}$\AA$^{-1}$s$^{-1}$sr$^{-1}$) in the EUV computed from the lower model component (bottom panel) and the upper model component (top panel). The lower components include lower layers of the atmosphere up to 10$^5$ K, the upper components include the upper TR and corona.
			\label{fig:euv}}
	\end{figure}

\begin{figure}
		\gridline{\fig{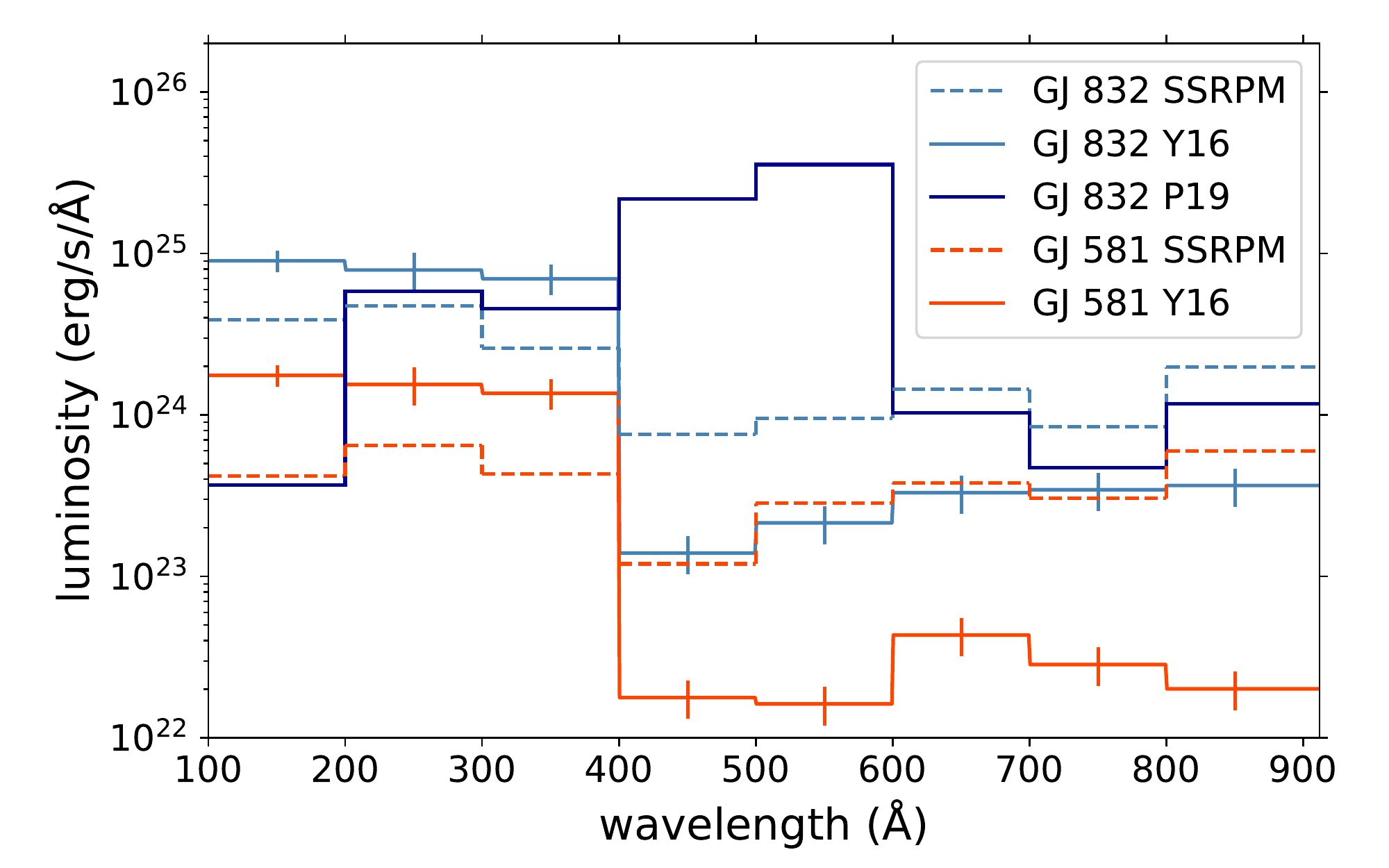}{0.5\textwidth}{}
			\vspace*{-20pt}
		}
		\caption{ Comparison between EUV luminosity estimates by \citet{lit:Youngblood2016} (denoted as Y16), the binned synthesized \texttt{PHOENIX} spectra from \citet{lit:Peacock832} (P19), and our binned EUV spectra. The vertical lines represent errors in the Y16 EUV reconstruction procedure.
			\label{fig:euv_c}}
	\end{figure}

	EUV spectra are shown in Figure \ref{fig:euv}, where we display contributions from lower ($T< 10^5$~K) and upper ($T>10^5$~K) component models separately. The integrated EUV luminosities between 100--912 \AA\ in our models are $\log{L_{\text{EUV},832}} = 27.25$ erg~s$^{-1}$ and $\log{L_{\text{EUV},581}} = 26.53$ erg~s$^{-1}$ (see Table \ref{tab:XrayEUV}). \citet{lit:Youngblood2016} used Ly$\alpha$-EUV empirical relations \citep{lit:LinskyLyalpha} to obtain EUV luminosities of these stars; their estimates -- $\log{L_{\text{EUV},832}} = 27.40$ erg~s$^{-1}$ and $\log{L_{\text{EUV},581}} = 26.67$ erg~s$^{-1}$ -- are very close to ours. However, we find that their SEDs in this wavelength range differ from ours by up to an order of magnitude, especially in the case of GJ 581 (Figure \ref{fig:euv_c}). We find that the lower components of model atmospheres contribute 24.1\% ($\log{L} = 26.63$ erg~s$^{-1}$) and 33.3\% ($\log{L} = 26.05$ erg~s$^{-1}$) of the EUV flux in GJ 832 and GJ 581, respectively. For GJ 581, we find that the flux below 400 \AA\ is underestimated, whereas above 400 \AA, the intensity is higher than that predicted by the empirical relation. A similar trend is seen in synthetic spectra of GJ 832 but to a lesser extent. The upper component model of GJ 581 contributes a significant fraction of flux above 400 \AA, which is not the case for GJ 832. Additionally, the discrepancy could be caused by the errors in Ly$\alpha$ reconstruction procedure, which could propagate and cause errors in reconstructed EUV fluxes.
	
\begin{deluxetable*}{lcccc}
	\centerwidetable
	\tablecaption{X-ray ($<100$ \AA) and EUV (100--912 \AA) Luminosities [erg s$^{-1}$]\label{tab:XrayEUV}}
	\tablehead{
	%\vspace{-5pt}
	    \colhead{} & \multicolumn{2}{c}{GJ 832} &  \multicolumn{2}{c}{GJ 581}  \\
		\colhead{} & \colhead{$\log L_{\text{X}}$}  & \colhead{$\log L_{\text{EUV}}$} & \colhead{$\log L_{\text{X}}$} & \colhead{$\log L_{\text{EUV}}$}
	}
	\startdata
	Present work & 26.73 &  27.25 & 25.93 & 26.53 \\
	\citet{lit:Fontenla832} & 26.73 &  27.26 & ... & ... \\
	\citet{lit:Loyd2016} & 26.26 & ...  & 25.98 & ... \\
	\citet{lit:Youngblood2016} & ... & 27.40 & ... & 26.67  \\
	\citet{lit:Peacock832} & ... & ~$27.85^a$ & ... & ... \\
	Duvvuri et al. (submitted) & ... & 26.83 & ... & ... \\
	\citet{lit:2011Sanz}$^b$~~ & 26.78 & 27.83 & ... & 27.18$^c$ \\
	\hline
	  & \multicolumn{4}{c}{$\log L(\text{90--360 \AA})$} \\
	Present work & \multicolumn{2}{c}{27.04} & \multicolumn{2}{c}{26.16}  \\
	\citet{lit:2018France} & \multicolumn{2}{c}{26.92} & \multicolumn{2}{c}{26.31} \\
	\enddata
	\vspace{5pt}
	\tablenotetext{\footnotesize{\textnormal{a}}}{ this is the revised value -- see discussion in Section \ref{sec:disc_euv}}
	\vspace{-5pt}
	\tablenotetext{\footnotesize{\textnormal{b}}}{ in their work, the X-ray range spans 5--100 \AA\ }
	\vspace{-5pt}
	\tablenotetext{\footnotesize{\textnormal{c}}}{ GJ 581 was not analyzed in the original work, but we used their scaling relation and our computed X-ray fluxes}
	
\end{deluxetable*}

	\citet{lit:2011Sanz} described an empirical relation between EUV and X-ray fluxes and estimated EUV luminosities for a series of stars, including GJ 832. Their estimated EUV flux, $\log{L_{\text{EUV},832}} = 27.83$ erg~s$^{-1}$ (Table \ref{tab:XrayEUV}), exceeds both the one reported in \citet{lit:Youngblood2016} and the one in the present work. However, as discussed in Section \ref{sec:spectra_xray}, they used a higher value of X-ray flux, which also increases EUV intensity. Using the computed X-ray flux from our model and their scaling relation, we obtain $\log{L_{\text{EUV},832}} = 27.71$ erg~s$^{-1}$ and $\log{L_{\text{EUV},581}} = 27.18$. While our computed EUV luminosities are lower, we note that they are well within the margin of error of these estimates \citep[see Eq. (3) in][]{lit:2011Sanz}. Additionally, the wavelength ranges they considered are different from ours: X-rays in their work extend from 5--100 \AA\ (as opposed to 1--100 \AA\ in the present work), and EUV covers an additional wavelength band between 912--920 \AA. Accounting for these differences could improve the agreement somewhat. Another empirical relation, based on FUV line fluxes, was described in \citet{lit:2018France} with the use of archival \textit{EUVE} data. They managed to estimate the fluxes of both GJ 832 and GJ 581 in the 90--360 \AA\ range, and these estimates agree reasonably well with our calculations (Table \ref{tab:XrayEUV}).
	
	The upcoming work by Duvvuri et al. (submitted) describes a DEM formalism and applies it to a series of stars, including GJ 832. They obtain $\log{L_{\text{EUV},832}} = 26.83$ erg~s$^{-1}$, but their DEM method does not take into account flux contributions due to various continua, of which the most prominent one, the H I Lyman continuum, contributes 14--15\% ($\log{L_{832}} = 26.42$ erg~s$^{-1}$ and $\log{L_{581}} = 25.67$ erg~s$^{-1}$) of the total EUV flux in our spectra. This is similar to what \citet{lit:2009Woods} found while analyzing Solar Irradiance Reference Spectra (SIRS), where this fraction is also 15\%.
	
\begin{figure}
		\gridline{\fig{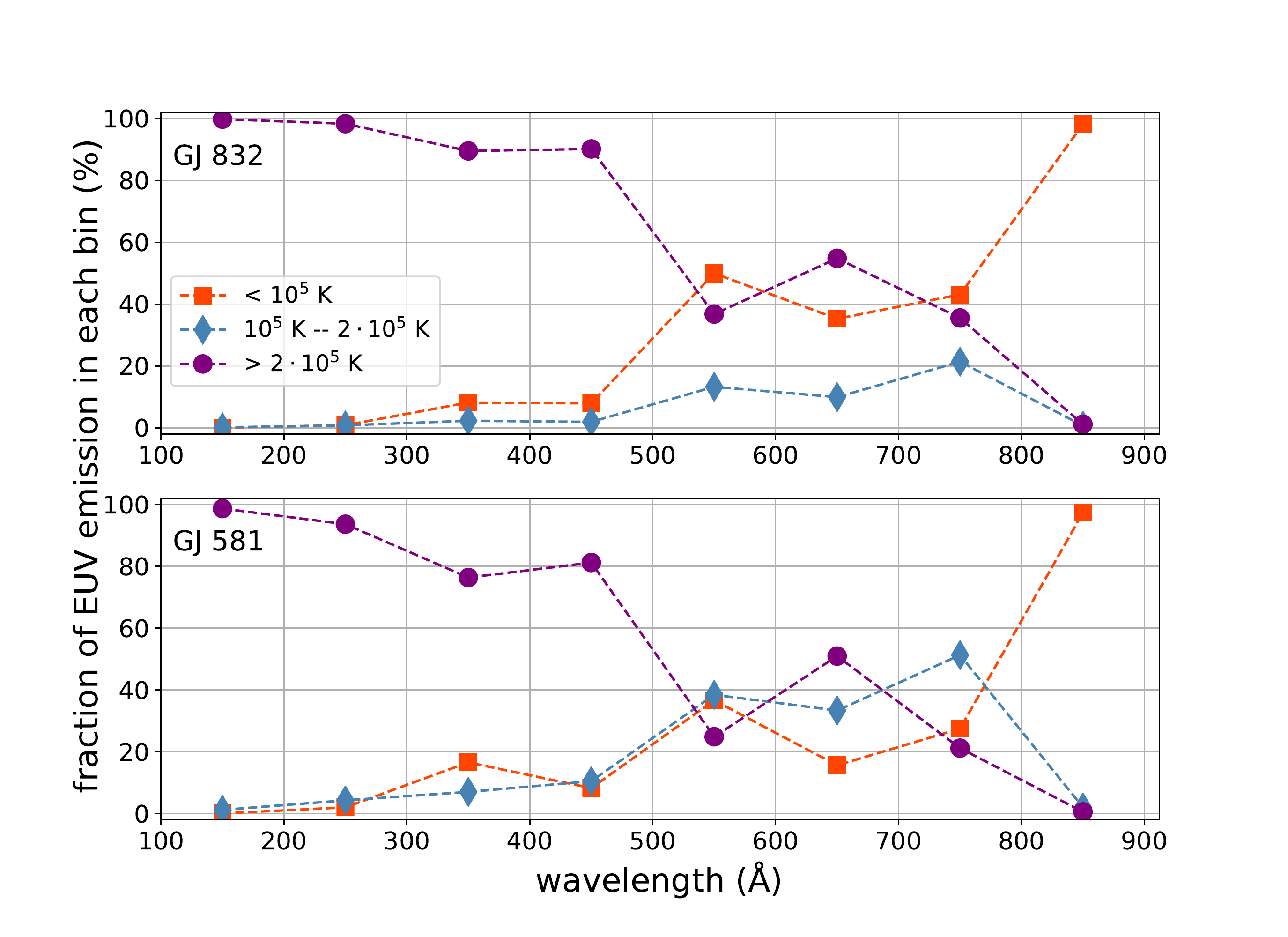}{0.5\textwidth}{}
			\vspace*{-20pt}
		}
		\caption{ Relative fractions of EUV emission from each component model in 100 \AA\ bins. The rightmost bin spans the range from 800--912 \AA. Temperatures below $<10^5$ K include the chromosphere and lower TR, the middle temperature range ($10^5-2\cdot10^5$ K) corresponds to the TR, and temperatures above $2\cdot10^5$ K correspond to the corona.
			\label{fig:euv_c_bymodel}}
	\end{figure}

	The atmosphere of GJ 832 has been approximated by one-dimensional models before. \citet{lit:Fontenla832}, whose approach we revise in the present work, used \texttt{SSRPM} to synthesize panchromatic spectra of this star, and their integrated X-ray and EUV fluxes are virtually indistinguishable from ours (Table \ref{tab:XrayEUV}). This is expected, as the primary difference between their model and ours lies in the photosphere and lower chromosphere, which do not significantly affect the short-wavelength emission. More recently, \citet{lit:Peacock832} synthesized the EUV spectra of GJ 832 using a one-dimensional \texttt{PHOENIX} model that did not include temperatures above 200,000 K. They argued that the omission of corona only significantly affects EUV emission below 200 \AA. In order to test whether our models support this claim, we built a separate model for each star that extends from $T=10^5$ K to $T=2\cdot10^5$ K and computed EUV spectra emerging from this layer. We find that the EUV contribution from temperatures above $2\cdot10^5$ K is 72.6\% ($\log{L_{832}} = 27.11$ erg~s$^{-1}$) and 52.4\% ($\log{L_{581}} = 26.24$ erg~s$^{-1}$) in GJ 832 and GJ 581, respectively. Fractions of EUV emission from the three components in 100 \AA\ bins are shown in Figure \ref{fig:euv_c_bymodel}. In our models, coronal EUV emission is significant, especially below 500 \AA. We caution, however, that splitting upper atmosphere models into separate components can produce erroneous results, as population densities at different heights are coupled in these highly non-LTE environments. \citet{lit:Peacock832} originally reported the integrated EUV flux of $\log{L_{\text{EUV},832}} = 27.31$ erg~s$^{-1}$, but this value was later revised and is now $\log{L_{\text{EUV},832}} = 27.85$ erg~s$^{-1}$ (Peacock 2020, personal communication). This value is very close to that reported in \citet{lit:2011Sanz} (Table \ref{tab:XrayEUV}). Their integrated EUV flux is higher than in the present work, and would be significantly higher when coronal emission is included. Their obtained SED is intermediate between ours and that reconstructed by \citet{lit:Youngblood2016} in most bands, except below 200 \AA, where the contribution from corona is significant, and between 400--600 \AA\ (Figure \ref{fig:euv_c}). Overall, the reasonably close agreement between different EUV reconstruction methods inspires confidence in each of them, but it is important to also accurately characterize the SED in the EUV range. 
	
	\subsection{Comparison of the Atmospheric Structures of GJ 832 and GJ 581} \label{sec:disc_rloss}
	
	GJ 832 and GJ 581 have similar photospheres (Figure \ref{fig:models}), which explains why their visible spectra are also comparable. The difference in lower photospheric temperature gradients is not significant; we find that it does not affect visible spectra. The lower chromospheres of the two stars are different, however: GJ 581 has a lower temperature minimum (2,300 K as opposed to 2,650 K for GJ 832), which is constrained by the K$_1$ flux of Ca II H \& K and the k$_1$ flux of the Mg II h \& k doublets.
	
	The first chromospheric rise is significantly less steep in the model of GJ 581 and occurs at lower pressures. While the pressure is relatively easy to infer using NUV spectral diagnostics, the steepness of the chromospheric rise is somewhat poorly constrained due to very low fluxes of potential diagnostic lines. Si II (1524 \AA) is a strong line that can be used as a diagnostic for this region in the solar atmosphere \citep{lit:1981Vernazza}, but we find that in the atmospheres of GJ 832 and GJ 581, the Si II doublet forms in the upper chromosphere between 15,000 and 20,000 K. The first chromospheric rise may be responsible for FUV continuum formation \citep{lit:SRPM2015}, although the temperature minimum region may also play a role \citep{lit:2010aHoudebine, lit:FAL2014}.
	
	The upper chromosphere and lower TR are very well constrained by bright NUV and FUV emission lines. The base of the TR is at a slightly lower pressure in GJ 581 (0.686 dyne cm$^{-2}$) than in GJ 832 (0.836 dyne cm$^{-2}$), and the TR itself is less steep -- at 100,000 K, the pressure is 0.682 dyne cm$^{-2}$ in GJ 581 and 0.835 dyne cm$^{-2}$ in GJ 832. A steeper TR of GJ 832 could indicate stronger magnetic heating. Other indicators of higher magnetic activity in GJ 832 compared to GJ 581 include brighter chromospheric and TR lines, such as Ca II, Mg II, Ly$\alpha$, and N V \citep{lit:Youngblood2017}, and more rapid rotation \citep[e.g., ][]{lit:2015West}. The activity indicator $\log R'_{HK}$ of GJ 832 exceeds that of GJ 581 by $\sim0.6$, which also suggests that GJ 832 is the more active star \citep{lit:2020Melbourne,lit:2017AsDe}.
	
	The coronal models are also qualitatively similar: they are both characterized by temperatures and pressures that are essentially constant with height (Figure \ref{fig:models}). The corona of GJ 581 is hotter than that of GJ 832, which may explain the possible excess EUV emission above 400 \AA. The coronal temperature in our model of GJ 581 also exceeds previous estimates.
	
	\subsection{Future Work} \label{sec:disc_fwork}
	
	There are several challenges to synthesizing accurate stellar spectra using \texttt{SSRPM}. Firstly, the atomic and molecular data should be updated. The versions of CHIANTI and NIST databases used in the present work are not fully up to date, which may cause inaccuracies in our computed spectra. More importantly, our databases lack some molecular opacity data, e.g., CO and H$_2$ opacities. H$_2$ is a major opacity source in the NUV, and H$_2$ populations are currently computed in chemical equilibrium. The agreement between computed and observed fluxes between 4000--4500 \AA\ could be improved by adding VO, CaH, and FeH molecular transitions. The missing UV opacity sources may result in enhanced ionization of neutral heavy species, which in turn leads to further decrease of UV opacity \citep{lit:SRPM2015}. So long as the photodissociation opacity of H$_2$ remains an \textit{ad hoc} parameter, the fitting of UV fluxes will remain an imprecise procedure. Additional opacity sources might alter the Ly$\alpha$ wings and the region between 4000 and 4500 \AA. The bound-free opacity edge at 1350 \AA, which produces abnormally high fluxes between 1350--1450 \AA, must be corrected in order for us to study the importance of FUV continuum. Atomic abundances used in the present work are identical to solar values, but until more precise and reliable measurements of stellar metallicities are available, this will continue to be a potential source of error.
	
	Another avenue to improve the fidelity of synthesized spectra is to build multi-component models for each star to account for inhomogeneity across stellar surfaces. M dwarfs are known to have active regions and starspots \citep[e.g., ][]{lit:2019Howard}, and with the recent launches of \textit{TESS} and \textit{Gaia}, it is becoming increasingly feasible to use photometry for estimating starspot coverage. Computing weighted averages of stellar surface fluxes should improve the fits between computed and observed spectra, inform our understanding of stellar thermal structures (especially in the chromosphere and corona), and provide an additional empirical constraint. However, this approach will still not account for important 3-D effects occurring at the boundaries of starspots, such as enhanced downward Ly$\alpha$ flux due to Wilson depressions \citep{lit:SRPM2015}. This particular effect can increase H$_2$ fluorescence, ionize heavier elements, and raise core fluxes for such lines as Mg II h \& k, Ca II H \& K, and Ly$\alpha$.
	
	Coronal emission is currently poorly constrained in our models, as we can only rely on a handful of faint emission lines and estimates of coronal temperatures. This introduces the possibility of degeneracy in T-P profiles of the upper atmospheres. We have not yet quantitatively studied the extent to which different coronal structures affect EUV emission, as it is currently computationally expensive to synthesize X-Ray and EUV spectra via \texttt{SSRPM}. Considering that most of the EUV emission is produced in the upper TR and the corona, it is important to have an accurate model of the corona for each target star. Further observations of the stars, especially if they are concurrent with UV observations, could improve model accuracy significantly. This would be especially useful for refining the coronal structure of GJ 581. Future EUV missions, such as the proposed \textit{ESCAPE}, will be invaluable, as presently there are virtually no empirical constraints on EUV fluxes of M dwarfs. Addtionally, it is important to constrain elemental abundances in stellar coronae to accurately predict coronal emission. For instance, when GJ 832 and GJ 581 coronal emissions were computed assuming solar coronal abundances \citep{lit:1992Feldman}, in which the fraction of iron is almost five times what it is in the photosphere due to first ionization potential (FIP) effect, the X-ray emission increased by a factor of 3-4. This is unlikely to be realistic, as M dwarf coronae are expected to have the inverse FIP effect \citep{lit:2012Wood}, but it shows that coronal emission is sensitive to elemental abundances in the corona.
	
	\section{Summary} \label{sec:outro}	
	
	We have built semi-empirical one-dimensional models for GJ 832 and GJ 581 using visible and UV spectra and X-ray fluxes. We used the full-NLTE radiative transfer code \texttt{SSRPM} that is equipped with an extensive database of atomic and molecular data. The synthetic spectra are in good agreement with observations, providing confidence in the estimates of unobservable UV fluxes. The most important results are:
	\begin{enumerate}
		\item Integrated extreme-UV fluxes of both stars were computed and are in reasonable agreement with estimates obtained using other methods, although the SEDs differ by up to an order of magnitude. In our models, upwards of 2/3 of the EUV total flux is formed at temperatures exceeding 10$^5$ K. EUV radiation is an essential input for describing photochemistry of planetary atmospheres and atmospheric mass loss, yet it is largely unobservable due to interstellar absorption. The outputs of our models can therefore be applied to exoplanetary studies.
		\item The far-UV continuum in our models is a major (42--54\%) contributor to overall FUV flux between 1450--1700 \AA, despite being dominated by noise in observations. While FUV continuum formation needs to be investigated further, this new result, if confirmed, has significant implications for photochemistry in exoplanetary atmospheres, since FUV radiation is responsible for photodissociating a number of important molecules.
		\item A comparison of atmospheric profiles of GJ 832 and GJ 581 reveals steeper chromospheric and TR temperature gradients, which may point to higher activity levels of GJ 832 compared to GJ 581. Higher fluxes of Mg II, Ca II, and a number of other lines that are proxies for magnetic activity corroborate this conclusion. So does the difference in computed FUV continuum brightness temperatures.
	\end{enumerate}
	
	A few specific ways to improve the models would be to update the atomic and molecular data and to build multi-component models for each star. While some inferences about the stellar structure can be made on the basis of the two modeled stars, more stellar models are needed in order to test our conclusions. It would be particularly insightful to compare the stellar models of the relatively inactive GJ 832 and GJ 581 with the more active M dwarfs of similar spectral types, such as GJ 644B, GJ 588, GJ 887, or GJ 205. Constraining the coronal structure also remains a challenge, and is particularly important for making robust EUV estimates.
	
	A more ambitious avenue for building accurate stellar models would involve NLTE formation for the most important molecular species, and spatial and temporal variability.
	
	\textit{ }
	
	\textbf{Data Availability:} The synthesized panchromatic (1 \AA\ -- 100 $\mu$m) spectra of GJ 832 and GJ 581 will be made available on the MUSCLES survey archive.\footnote{\href{https://doi.org/10.17909/T9DG6F}{https://doi.org/10.17909/T9DG6F}}
	
	\textit{ }
	
	\textbf{Acknowledgements:} Based on data acquired at Complejo Astronómico El Leoncito, operated under agreement between the Consejo Nacional de Investigaciones Científicas y Técnicas de la República Argentina and the National Universities of La Plata, Córdoba and San Juan. The authors thank the MUSCLES and HAZMAT collaborations for making their spectral data publicly available. The authors also thank the referee for the many useful comments and suggestions. The radiative transfer code \texttt{SSRPM} was developed by Dr. Juan Fontenla, who guided this research program until his untimely death in January 2018. This multiyear theory program is supported by grant HST-AR-15038.001 from the Space Telescope Science Institute, which is operated by the Association of Universities for Research in Astronomy, Inc. for NASA, under contract NAS 5-26555.
	
	\textit{ }
	
	\textbf{Software:} \texttt{SRPM / SSRPM} \citep[][and references within]{lit:Fontenla832,lit:SRPM2015}, \texttt{astropy} \citep{soft:astropy}, \texttt{numpy} \citep{soft:numpy}, \texttt{matplotlib} \citep{soft:plt}

	\clearpage

	\appendix
    \restartappendixnumbering
    
    \begin{deluxetable}{lchccccccc}[b]
		\tablenum{A1}
		\vspace{10pt}
		\tablecaption{Elemental Abundances and Full-NLTE Parameters \label{tab:abunslvls}}
		\tablewidth{0pt}
		\tablehead{
		    %\vspace{-5pt}
			\colhead{} & \colhead{Levels/} & \nocolhead{} & \colhead{} & \colhead{Levels/} & \colhead{} & \colhead{Levels/} & \colhead{} \\
			\colhead{Ion} & \colhead{Subevels} & \nocolhead{} & \colhead{Ion} & \colhead{Sublevels} & \colhead{Ion} & \colhead{Sublevels} & \colhead{$n_i/n_H$}
		}
		\vspace{-10pt}
		\startdata
		H I & 15/15 & nada & ... & ... & ... & ... & 1.0 \\
		He I & 20/32 & nada & He II & 15/25 & ... & ... & 0.1 \\
		C I & 45/87 & nada & C II & 27/50 & ... & ... & 2.4E-4 \\
		N I & 26/61 & nada & N II & 33/61 & ... & ... & 9.0E-5 \\
		O I & 23/51 & nada & O II & 31/68 & ... & ... & 3.9E-4 \\
		Ne I & 80/80 & nada & Ne II & 57/57 & ... & ... & 6.9E-5 \\
		Na I & 22/29 & nada & Na II & 14/25 & ... & ... & 3.0E-6 \\
		Mg I & 26/44 & nada & Mg II & 14/23 & Mg III & 54/93 & 3.4E-5 \\
		Al I & 18/31 & nada & Al II & 20/34 & Al III & 32/42 & 4.0E-6 \\
		Si I & 35/65 & nada & Si II & 14/25 & Si III & 60/108 & 3.2E-5 \\
		S I & 20/50 & nada & S II & 30/65 & ... & ... & 6.9E-6 \\
		Ar I & 48/48 & nada & Ar II & 57/57 & ... & ... & 1.5E-6 \\
		K I & 10/16 & nada & ... & ... & ... & ... & 2.5E-7 \\
		Ca I & 22/38 & nada & Ca II & 24/33 & Ca III & 34/65 & 2.0E-6 \\
		Ti I & 80/204 & nada & Ti II & 78/204 & Ti III & 43/83 & 7.9E-8 \\
		V I & 80/240 & nada & V II & 41/111 & V III & 40/99 & 1.0E-8 \\
		Cr I & 80/265 & nada & Cr II & 34/95 & Cr III & 20/50 & 4.4E-7 \\
		Mn I & 85/261 & nada & Mn II & 28/74 & Mn III & 40/112 & 2.5E-7 \\
		Fe I & 80/253 & nada & Fe II & 80/221 & Fe III & 80/213 & 2.8E-5 \\
		Co I & 65/167 & nada & Co II & 28/77 & Co III & 50/143 & 8.3E-8 \\
		Ni I & 61/136 & nada & Ni II & 28/68 & Ni III & 40/102 & 1.7E-6 \\
		\enddata
		
	\end{deluxetable}
    
    Atomic and molecular level populations in each model are initialized in LTE. This assumption is valid for the majority of species in the photosphere, where plasma density is sufficiently high for collisions to dominate. Non-zero departures from LTE are, however, relevant in the photosphere \citep[e.g.,][]{lit:nickellines,lit:2014Mashonkina}, hence it is included in the more complex computations described below. In the outer layers, density decreases rapidly, and non-radiative heating raises temperatures well above the radiative-convective equilibrium predictions. Due to the optically thin environment of the upper TR and corona, we compute level populations for highly ionized atoms in the effectively optically thin NLTE approximation. This approximation is appropriate because the spectral lines associated with transitions between energy levels of these species have optical thickness that is exceeded by the scattering length scales. The selected atoms and ions (see Table \ref{tab:abunslvls}) which are abundant enough to cause significant scattering in their associated spectral lines are computed in full-NLTE. The NLTE computations are performed in iteration blocks. Within each iteration, the equations for statistical equilibrium are represented as a system of non-linear equations of the order $m\cdot n$, where $m$ is the number of levels and $n$ is the number of height points. The iterative procedure solves for radiative and collisional transitions using the values of previous iterations. Additionally, the procedure contains a linearization scheme, allowing for the system of equations to be linearized and for the transitional values to be solved for from the linearized system. The values obtained in linearized and non-linearized computations are compared after each iteration and recomputed until they converge to within 5--10\%. The number of required iterations is typically between 10 and 50 and is given as an input for the program in configuration files, along with several other numerical and physical parameters. These parameters include but are not limited to: parameters required for computation of the effect of ambipolar diffusion, parameters pertaining to numerical stability of the code, and binary parameters that determine e.g. whether to use CRD or PRD and whether to compute pressures at different grid points using the hydrostatic equilibrium. Since the plasma densities in the upper TR and the corona are too low for optically thick lines to form, we split the atmosphere models into two components: one that extends from the photosphere to $10^5$ K, and the other that extends from $10^5$ K to the top of the corona. Accordingly, for the second component, we compute all atoms and ions in the optically thin approximation, while for the first component, the highly ionized species are not included in computations, since temperatures do not exceed $10^5$ K. Both adjustments are made to significantly reduce computing time. The interface between the two components is computed by matching the boundary conditions.
    
    The emerging spectra are computed in terms of surface flux, or average disk intensity, using plane-parallel symmetry for the lower component model and spherical symmetry for the upper TR and corona. The resolution of the synthetic spectra is $R=\lambda/\Delta\lambda \approx 10^6$. The conversion factor between the surface flux and the flux at Earth is $\Omega = \pi (R/d)^2$ In order to compare models with observations, the observed spectra are converted from flux at Earth to stellar surface flux. That is, distances to the stars of interest and their radii are used in the conversion procedure. As can be seen from Table \ref{tab:stellarparams}, this introduces some error, especially for the radii as their associated errors are considerably larger than the errors in parallax measurements. Once synthetic spectra are smoothed to match the resolution of the instrument used to observe the star, we compare the spectra and note wavelengths and spectral features where the fit is poor. We then compute the contribution function and the optical depth as functions of height for these features and adjust the atmospheric structure accordingly.
	
    %\clearpage

		\bibliography{sample63}{}
		
		\bibliographystyle{aasjournal}
		
	\end{document}